\documentclass[twocolumn,journal]{IEEEtran}
\usepackage[T1]{fontenc}
\usepackage[latin9]{inputenc}
\usepackage{color}
\usepackage{array}
\usepackage{rotating}
\usepackage{float}
\usepackage{booktabs}
\usepackage{units}
\usepackage{bbding}
\usepackage{multirow}
\usepackage{amsmath}
\usepackage{amsthm}
\usepackage{amssymb}
\usepackage{undertilde}
\usepackage{graphicx}

\makeatletter

\newcommand\blfootnote[1]{%
  \begingroup
  \renewcommand\thefootnote{}\footnote{#1}%
  \addtocounter{footnote}{-1}%
  \endgroup
}

\newcommand{\noun}[1]{\textsc{#1}}
\providecommand{\tabularnewline}{\\}
\floatstyle{ruled}
\newfloat{algorithm}{tbp}{loa}
\providecommand{\algorithmname}{Algorithm}
\floatname{algorithm}{\protect\algorithmname}

\theoremstyle{plain}
\newtheorem{thm}{\protect\theoremname}
\theoremstyle{remark}
\newtheorem{rem}[thm]{\protect\remarkname}

\usepackage{algorithm}
\usepackage{cite}
\usepackage{xcolor}
\usepackage{booktabs}
\usepackage[noend]{algpseudocode}
\usepackage[caption=false,font=footnotesize]{subfig}

\@ifundefined{showcaptionsetup}{}{%
 \PassOptionsToPackage{caption=false}{subfig}}
\usepackage{subfig}
\makeatother

\providecommand{\remarkname}{Remark}
\providecommand{\theoremname}{Theorem}

\begin{document}
\title{Modeling, Simulating and Configuring Programmable Wireless Environments
for Multi-User Multi-Objective Networking}
\author{Christos Liaskos\IEEEauthorrefmark{1}, Ageliki Tsioliaridou\IEEEauthorrefmark{1},
Shuai Nie\IEEEauthorrefmark{3}, Andreas Pitsillides\IEEEauthorrefmark{2},
Sotiris Ioannidis\IEEEauthorrefmark{1}, and Ian Akyildiz\IEEEauthorrefmark{2}\IEEEauthorrefmark{3}\\
{\small{}\IEEEauthorrefmark{1}Foundation for Research and Technology
- Hellas (FORTH)}\\
{\small{}Emails: \{cliaskos,atsiolia,sotiris\}@ics.forth.gr}\\
{\small{}\IEEEauthorrefmark{2}University of Cyprus, Computer Science
Department}\\
{\small{}Email: Andreas.Pitsillides@ucy.ac.cy}\\
{\small{}\IEEEauthorrefmark{3}Georgia Institute of Technology, School
of Electrical and Computer Engineering}\\
{\small{}Email: ian@ece.gatech.edu}}
\maketitle
\begin{abstract}
Programmable wireless environments enable the software-defined propagation
of waves within them, yielding exceptional performance potential.
Several building-block technologies have been implemented and evaluated
at the physical layer. The present work contributes a network-layer
scheme to configure such environments for multiple users and objectives,
and for any physical-layer technology. Supported objectives include
any combination of Quality of Service and power transfer optimization,
eavesdropping and Doppler effect mitigation, in multi-cast or uni-cast
settings. Additionally, a graph-based model of programmable environments
is proposed, which incorporates core physical observations and efficiently
separates physical and networking concerns. Evaluation takes place
in a specially developed, free simulation tool, and in a variety of
environments. Performance gains over regular propagation are highlighted,
reaching important insights on the user capacity of programmable environments.
\end{abstract}


\IEEEpeerreviewmaketitle{}

\section{Introduction\label{sec:Introduction}}

\IEEEPARstart{R}{ecent} years have seen the rise of efforts to control
the wireless propagation within a space, introducing programmable
wireless environments (PWEs)~\cite{Liaskos:2018:UAS:3289258.3192336}.
According to the PWE paradigm, planar objects\textendash such as walls
in a floorplan\textendash receive a special coating that can sense
impinging waves and actively modify them by applying an electromagnetic
(EM) \emph{function}. Examples include altering the wave's direction,
power, polarization and phase~\cite{MSSurveyAllFunctionsAndTypes}.
The capabilities of several coating technologies have been demonstrated~\cite{Yang.2016,Tan.2018,Welkie.2017}.
The present work builds upon these physical-layer works, and proposes
a solution to the network-layer PWE configuration problem, i.e., which
functions to deploy at the PWE coatings to serve a set of given user
communication objectives.\blfootnote{This work is part of project VISORSURF: A HyperVisor for Metasurface Functionalities (www.visorsurf.eu). Funded by the European Union Horizon 2020, under the Future Emerging Technologies - Research and Innovation Actions call (Grant Agreement EU 736876).}

Evaluated coating technologies for PWEs include relays, phased antenna
arrays, and metasurfaces~\cite{Liaskos2019ADHOC}. Each technology
comes with a range of supported functions, environmental applicability
and efficiency degrees. Relays are 1~input-~N~output antenna pairs
that can be placed over walls at regular intervals~\cite{Welkie.2017}.
At each pair, one out of the N outputs can be selected, thereby redirecting
the input wave in a partially customizable manner. Phased antenna
arrays\textendash also known as intelligent surfaces and reflectarrays~\cite{RadarDiscrete,Subrt.2012,Tan.2016}\textendash are
panels commonly comprising a number of patch antennas with half-wavelength
size, in a 2D grid arrangement. At each patch, active elements such
as PIN diodes are used for altering the phase of the reflected EM
wave. Consistent wave steering and absorption is attained at the far
field. Metasurfaces are similar structures, but with a $25-100^{+}$
times higher density of \emph{meta-atoms} (i.e, the repeating unit
of a planar antenna and active elements)~\cite{MSSurveyAllFunctionsAndTypes}.
This density allows them to form any surface current distribution
over them, thereby producing any EM output due to the Huygens principle~\cite{Pao.1976}.
Thus, highly efficient EM functions even in the near field can be
attained. HyperSurfaces are a novel class of networked metasurfaces
that comes with a software programming interface (\emph{API}) and
an \emph{EM compiler}~\cite{LiaskosAPI,LiaskosComp}. The API allows
for getting the HyperSurface state and setting its EM function, while
abstracting the underlying physics. The EM compiler translates the
API callbacks to corresponding active element states.

A PWE is created by coating planar objects\textendash such as as walls
and ceilings in an indoor environment\textendash with \emph{tiles},
i.e., rectangular panels of any aforementioned technology, with inter-networking
capabilities~\cite{Liaskos:2018:UAS:3289258.3192336}. The latter
allow a central server to connect to any tile, get its state and set
its EM function in an automated manner~\cite{IEEEcomLiaskos}. This
maturity level reached at the physical layer of tiles opens a new
research direction at the network level: \emph{given a set of users
with communication objectives within a PWE, what is the optimal EM
function per tile to serve them}?

The present work contributes a solution to this problem, able to handle
multiple users, objectives and EM functions. User mobility, multiple
objectives per user, multicast groups and partially coated PWEs are
supported. The objectives include power transfer and signal-to-interference
maximization, as well as eavesdropping and Doppler effect mitigation.
In order to achieve these traits, the work also contributes:
\begin{itemize}
\item A systematic way of formulating and combining EM functions, which
takes into account key-outcomes from the field of Physics (metamaterials).
\item The \emph{EM profile} of tiles, a novel concept that describes the
supported EM functions per tile and their efficiency.
\item A graph-based model to describe PWEs, and a way of transforming communication
objectives to graph paths.
\item A novel tool specifically developed for realistically simulating PWEs.
\end{itemize}
Extensive evaluations in multiple floorplans and topologies yield
important conclusions about the maximum potential of PWEs and their
user capacity in terms of maximal supported traffic load.

The remainder of this work is organized as follows. Section~\ref{sec:Related-work}
surveys related studies. Section~\ref{sec:A-Graph-based-Model} describes
the graph-based modeling of PWEs and the concepts of tile EM functions
and profile. Section~\ref{sec:A-K-paths-Approach} details the novel
scheme for configuring PWEs. Evaluation takes place in Section~\ref{sec:Evaluation}.
The discussion follows in Section~\ref{sec:Discussion-and-Future},
along with future work directions, and the paper concludes in Section~\ref{sec:Conclusion}.

\section{Related work\label{sec:Related-work}}

PWEs are attracting attention due to the recent advances in the development
of new techniques to control the radiation patterns of EM waves~\cite{AccessUPC,Subrt.2012,DBLP:conf/WoWMoM/Liaskos,IEEEcomLiaskos,I.F.Akyildiz.2018}.
The existing literature mainly refers to PWE tile unit technologies,
rather than PWE configuration approaches. We employ the layered taxonomy
of Fig.~\ref{figArch} (introduced in~\cite{AccessUPC}) to survey
them:

\begin{figure}[t]
\begin{centering}
\includegraphics[viewport=0bp 0bp 853bp 482bp,width=1\columnwidth]{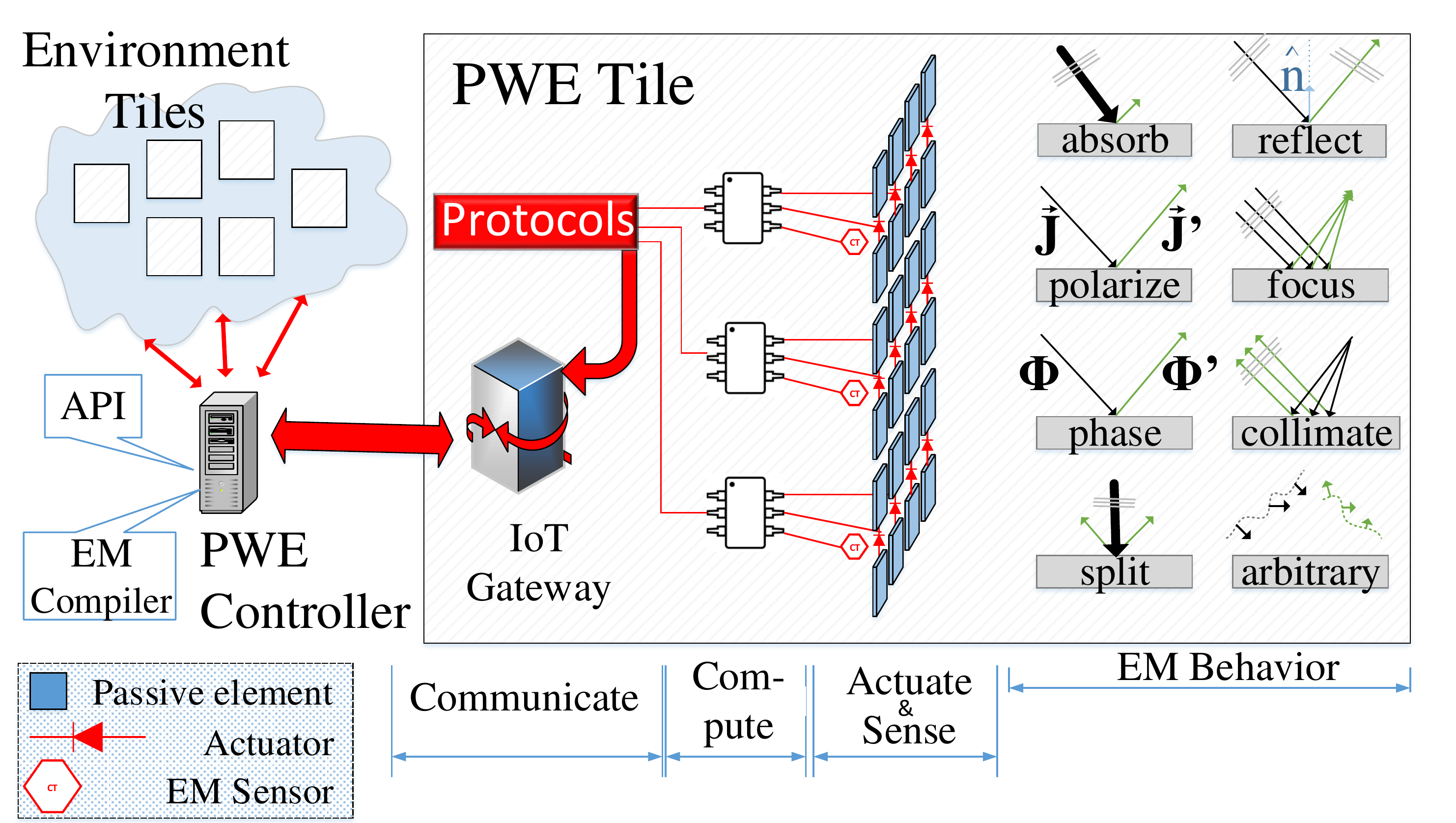}
\par\end{centering}
\caption{\label{figArch}Tile composition and PWE architecture. }
\end{figure}
\textbf{EM behavior Layer.} This layer comprises the supported EM
function of the tile, and its principle of operation. Reflectarray
tiles (and also phased arrays or intelligent surfaces) employ modifiable
phase shifts applied over their surface. At the far field, reflected
rays can be considered co-directional, and their superposition\textendash constructive
or destructive\textendash is controlled by the applied phase shifts~\cite{RadarDiscrete}.
Thus, wave scattering or redirection functions can be attained. Metamaterial
tiles operate at a lower level, acting as surfaces with tunable local
impedance~\cite{MSSurveyAllFunctionsAndTypes}. Impinging waves create
inductive surface currents over the tile, which can be routed by tuning
the local impedance across the tile. Notice that the principle of
Huygens states that any EM wavefront can be traced back to a current
distribution over a surface~\cite{Pao.1976}. Thus, in principle,
metamaterials can produce any custom EM function as a response to
an impinging wave. Common functions include wave steering, focusing,
collimating (i.e., producing a planar wavefront as a response to an
impinging wave), polarizing, phase altering, full or partial absorption,
frequency selective filtering and even modulation~\cite{MSSurveyAllFunctionsAndTypes,CUIcoding2018}.
Metamaterials can be classified further as non-plasmonic or plasmonic.
In the former, the impinging wave does not affect the configured local
impedance. In plasmonic metamaterials, the surface impedance is altered
by the impinging wave, producing non-linear effects~\cite{PlasmonicMS,Lockyear.2009}.
Thus, plasmonic metamaterials pose extra challenges in exerting deterministic
control over waves, and are not considered for PWEs in the present
work.

\textbf{Actuation and Sensing Layer.} This layer includes the actual
hardware elements that can be controlled to achieve a phase shift
or impedance distribution across a tile. Commonly, the layer comprises
arrays of planar antennas\textendash such as copper patches\textendash and
multi-state switches between them. Reflectarray tiles usually employ
PIN diodes with controllable biasing voltage as switches~\cite{Yang.2016}.
Metamaterials have employed a wider range of choices, both in the
shape and geometry of the planar antennas and in the nature of switches.
CMOS transistors, PIN diodes, Micro-Electro-Mechanical Switches (MEMS),
micro-fluidic switches, magnetic and thermal switches are but a few
of the considered options in the literature~\cite{Chen.2016}. Notably,
some options\textendash such as micro-fluid switches\textendash are
state-preserving in the sense that they require power only to change
state but not to maintain it (i.e., contrary to biased PIN diodes).

Sensing impinging waves is also necessary for exerting efficient control
over them. While this information can be provided by external systems~\cite{IEEEcomLiaskos},
tiles can incorporate sensing capabilities as well~\cite{NANOCOMlocalizeSDM}.
The sensing can be direct, employing specialized sensors~\cite{DDOT.b},
or indirect, e.g., by deducing some impinging wave attributes from
currents or voltages between tile elements~\cite{AccessUPC}.

\textbf{Computing Layer.} This layer comprises the computing hardware
that controls the actuating and sensing elements. Its minimum duties
include the mapping of local phase or impedance values to corresponding
actuator states. Reflectarray tiles commonly implement this layer
using FPGAs and shift registers~\cite{Yang.2016}. Metasurfaces,
and specifically HyperSurfaces, can alternatively employ standard
Internet-of-Things (IoT) devices for the same purpose~\cite{DBLP:conf/WoWMoM/Liaskos}.
Moreover, they can optionally include application-specific integrated
circuits (ASICs) distributed over the tile meta-atoms~\cite{pitilakisMETA2018,DBLP:conf/iscas/Tasolamprou}.
This can enable autonomous, ``thinking'' tiles, where meta-atoms
detect the presence and state of one another, and take local actuating
decisions to meet a general functionality objective. Nonetheless,
this advanced capabilities are not required for PWEs.

\textbf{Communication Layer.} This layer comprises the communication
stack and the means that connect the actuating and sensing layers,
the computing layer and tile-external devices (including other tiles
and computers that monitor and configure PWEs~\cite{Liaskos:2018:UAS:3289258.3192336}).
In the simplest case, this layer is implemented within the computing
hardware, acting as a gateway to the tile-external world, using any
common protocol (e.g., Ethernet). HyperSurface tiles with embedded
ASICs additionally require inter-tile communication schemes, to handle
the information exchange between smart meta-atoms. Both wired and
wireless intra-tile communication is possible~\cite{pitilakisMETA2018,DBLP:conf/iscas/Tasolamprou}.
In both cases, the ASIC hardware employs custom, non-standard protocols.

\emph{Differentiation}. The related studies focus on one or more tile
layers. However, to the best of the authors' knowledge, the topic
of configuring a PWE per user directives has not been previously studied.
In their previous work, the authors formally introduced the PWE concept,
its architecture and challenges~\cite{IEEEcomLiaskos,Liaskos:2018:UAS:3289258.3192336}.
Proof of concept PWE simulations took place in~\cite{DBLP:conf/WoWMoM/Liaskos,Liaskos2019ADHOC},
which treated the PWE configuration problem as a block box, employing
a genetic algorithm to configure the PWE tile functions. Power maximization
over an area served as the driving criterion for the genetic algorithm.
The present work departs from genetic heuristics and offers an exact
configuration process. The novel process can handle multiple users
and objectives spanning security, quality of service (QoS), mobility
and wireless power transfer. A graph-based model for PWEs is also
introduced, that can facilitate future contributions in the network-layer
of PWEs, using the related physical-layer studied in tile technologies
as input.

\section{A Graph-based Model for Programmable Environments\label{sec:A-Graph-based-Model}}

\textcolor{black}{This Section provides an abstract model of the Physics
behind metasurfaces, leading to a function-centric formulation of
their capabilities. This formulation is then used for modeling PWEs
as a graph, and describing its workflow and performance objectives
as path finding problems. }

\textcolor{black}{With no loss of generality, the text considers HyperSurface
tiles, since they offer the richest set of supported features. The
model, however, remains valid for any other tile technology. Moreover,
the study will use an indoors setting as the driving scenario, but
it remains valid in any other setting. Finally, the operating principles
of PWEs and metasurfaces described in Sections~\ref{sec:Introduction}
and~\ref{sec:Related-work} are sufficient for the remainder of the
text. Additional introductory material can be found in the literature~\cite{IEEEcomLiaskos,Liaskos:2018:UAS:3289258.3192336,Liaskos2019ADHOC}.}

Persistent notation is summarized in Table~\ref{tab:Summary-of-Notation}
for ease. (Notation used only locally in the text is omitted).
\begin{table}[t]
\begin{centering}
\caption{\noun{Summary of Notation\label{tab:Summary-of-Notation}}}
\par\end{centering}
\centering{}%
\begin{tabular*}{1\columnwidth}{@{\extracolsep{\fill}}|c|l|}
\hline
Symbol  & Explanation\tabularnewline
\hline
\hline
$\mathcal{H}$ & The set of all tiles within an environment.\tabularnewline
\hline
$h\in\mathcal{H}$  & A single HyperSurface tile.\tabularnewline
\hline
$\mathcal{F}_{h}$  & The set of EM functions supported by a tile $h$.\tabularnewline
\hline
$f_{h}\in\mathcal{F}_{h}$  & A single function, deployed to tile $h$.\tabularnewline
\hline
$f_{h}^{\text{\textsc{Abs}}},f_{h}^{\text{\textsc{Str}}},f_{h}^{\text{\textsc{Col}}}$  & Absorption, Steering and Collimation functions. \tabularnewline
\hline
$m_{h}^{\text{\textsc{Pha}}},m_{h}^{\text{\textsc{Pol}}}$  & EM phase and polarization function modifiers.\tabularnewline
\hline
$\overrightarrow{E_{in}}$,~$\overrightarrow{E_{out}}$  & Nominal input/output (EM field) of a function.\tabularnewline
\hline
$\mathbb{I}$,~$\mathbb{O}:$  & EM function input/outputs as wave attributes:\tabularnewline
$\left\langle \omega,\,\overline{D},\,\mathbf{P},\,\overrightarrow{\mathbf{J}},\,\Phi\right\rangle $ & <frequency, direction, power, polarity, phase>.\tabularnewline
\hline
$_{PW},\,_{FW}$ & Subscripts denoting plain wave and focal wave.\tabularnewline
\hline
$g_{h}$ & Wave power gain/loss after impinging at tile $h$.\tabularnewline
\hline
\multirow{2}{*}{ ${\scriptstyle \mathcal{G}\left\langle \left\{ \mathcal{H},\,\mathcal{U}\right\} ,\left\{ \mathcal{L}_{h},\,\mathcal{L}_{u}\right\} \right\rangle }$} & Graph with tiles $\mathcal{H}$ and users $u\in\mathcal{U}$ as nodes, \tabularnewline
 & inter tile links $\mathcal{L}_{h}$ and user-to-tile links $\mathcal{L}_{u}$.\tabularnewline
\hline
$\overrightarrow{p}_{n\,n'}$ & A path in $\mathcal{G}$ as list of links from node $n$ to $n'$.\tabularnewline
\hline
$l_{n\,n'}$ & A link in $\mathcal{G}$ from node $n$ to $n'$.\tabularnewline
\hline
$\text{\textsc{Tx}}\left(l_{uh}\right),\text{\textsc{Rx}}\left(l_{hu}\right)$ & Link labels denoting intended Tx and Rx users.\tabularnewline
\hline
$\left\langle \ldots\right\rangle $ & A tupple (group) of items.\tabularnewline
\hline
$\left\{ \ldots\right\} $ & A list of objects (single items or tupples).\tabularnewline
\hline
$\widetilde{\phantom{E}*\phantom{E}}$ & Unintended (not nominal) type of quantity $*$. \tabularnewline
\hline
$\left\Vert *\right\Vert $  & The cardinality of a set $*$.\tabularnewline
\hline
\end{tabular*}
\end{table}

\subsection{General Modeling and Properties of of HyperSurface Functions}

\begin{figure}[t]
\begin{centering}
\includegraphics[viewport=0bp 220bp 730bp 550bp,width=1\columnwidth]{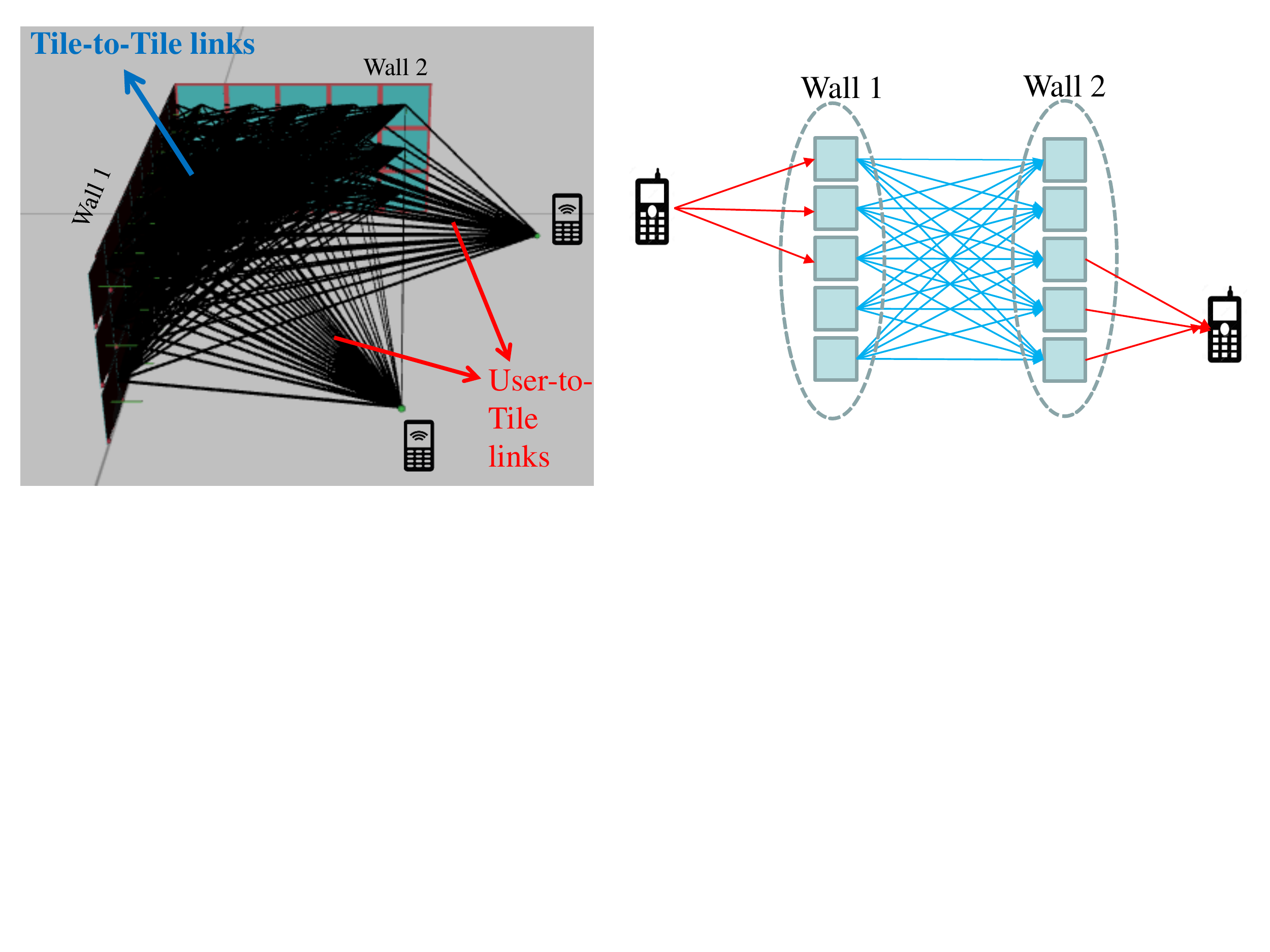}
\par\end{centering}
\caption{\label{fig1}3D and 2D illustration of the types of links (user links
and inter-tile links) and nodes (tiles and users) in a PWE. }
\end{figure}

Let $\mathcal{H}$ denote the set of all HyperSurface tiles deployed
within an environment, such as the floorplan of Fig.~\ref{fig1}.
A single tile will be denoted as $h\in\mathcal{H}$. Let $\mathcal{F}_{h}$
denote all possible EM functions that can be deployed to a tile $h$.
A single function deployed to a tile will be $f_{h}\in\mathcal{F}_{h}$.

As discussed in Sections \ref{sec:Introduction} and~\ref{sec:Related-work},
a function $f_{h}$ is attained by setting the active elements of
the HyperSurface accordingly. In this work we will assume that the
correspondence between functions and active element states is known,
and the reader is redirected to studies on \emph{EM Compilers} for
further details~\cite{LiaskosAPI,LiaskosComp}.

\begin{figure}[t]
\begin{centering}
\includegraphics[viewport=0bp 0bp 450bp 250bp,clip,width=0.7\columnwidth]{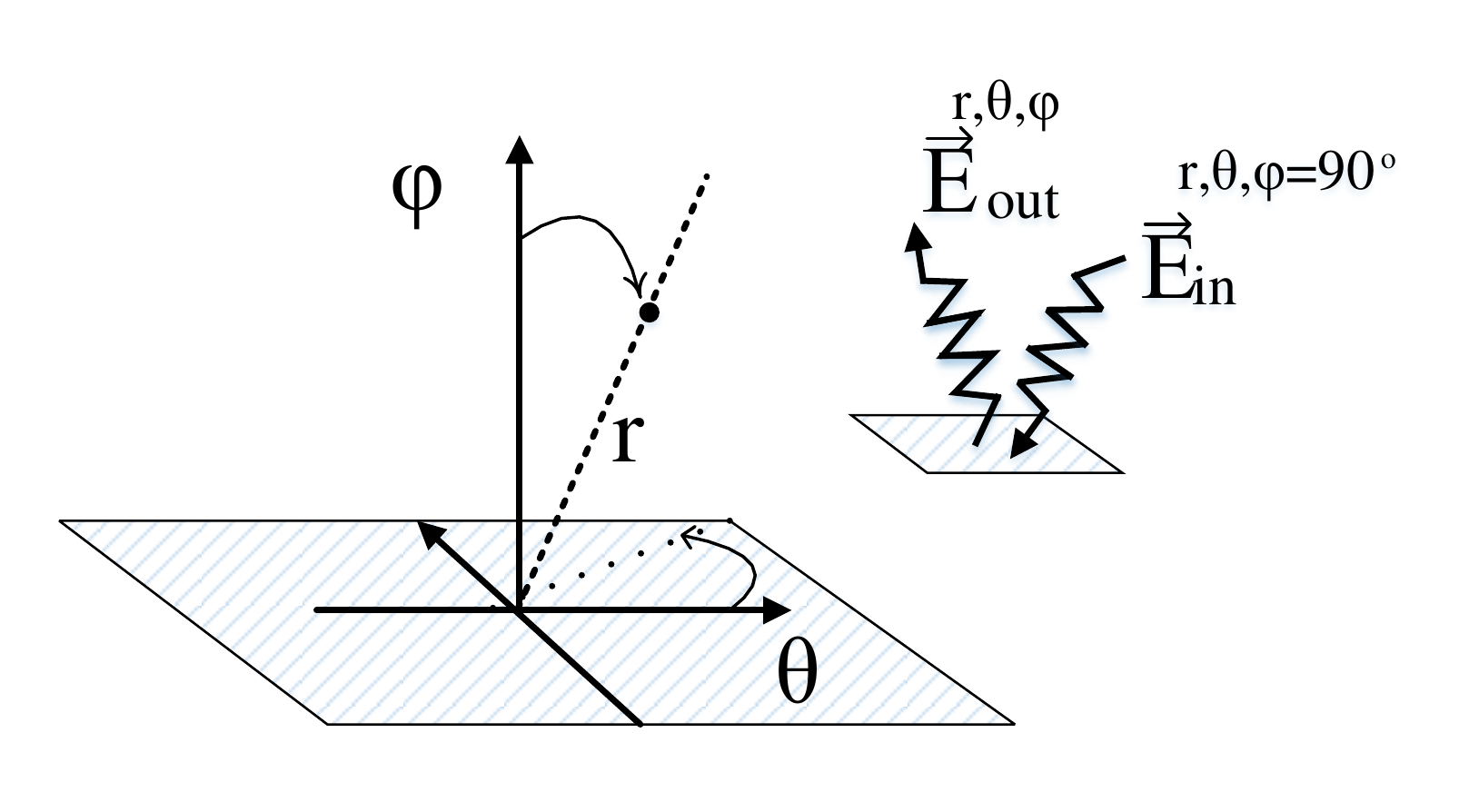}
\par\end{centering}
\caption{\label{fig2}The tile coordinate system for describing its inputs
and outputs. The origin is at the tile center.}
\end{figure}
Each function $f_{h}$ receives a nominal input EM field, $\overrightarrow{E_{in}}$,
(i.e., impinging upon the tile), and then returns a well-defined output
$\overrightarrow{E_{out}}$ (i.e., a reflected, refracted or no field\textendash in
case of perfect absorption), which can be abstracted as:

\begin{equation}
\overrightarrow{E_{out}}\gets f_{h}\left(\overrightarrow{E_{in}}\right)\label{eq:1}
\end{equation}
Consider the coordinate system over a tile, as shown in Fig.~\ref{fig2}.
In the most generic function case, $\overrightarrow{E_{in}}$ is defined
over the $\phi=90^{o}$ plane on the surface, while $\overrightarrow{E_{out}}$
contains the output field at any point $\left\{ r,\theta,\phi\right\} $
around the tile. It is noted that a function $f_{h}$ also defines
the output to any, even \emph{unintended input}, $\widetilde{E_{in}}$,
which can exemplary arise when EM sources move, without adapting the
tile functions accordingly. Therefore, relation (\ref{eq:1}) is generalized
as:

\begin{equation}
\widetilde{E_{out}}\gets f_{h}\left(\widetilde{E_{in}}\right)\label{eq:2}
\end{equation}
We proceed to remark two important properties of the EM functions,
stemming from physics:
\begin{rem}
\label{rem:F_symmetry-}EM functions $f_{h}$ are symmetric~\cite{MSSurveyAllFunctionsAndTypes,PlasmonicMS}:

\begin{equation}
\widetilde{E_{out}}\gets f_{h}\left(\widetilde{E_{in}}\right)\Leftrightarrow\widetilde{E_{in}}\gets f_{h}\left(\widetilde{E_{out}}\right)\label{eq:symmetry}
\end{equation}
\end{rem}
The symmetry remark can be used for defining a common format for inputs
and outputs in Section~\ref{subsec:modelIO}. It will also be called
upon later on, to ensure that communication channels created by tuning
HyperSurfaces are bidirectional.
\begin{rem}
\label{rem:F_linearity}EM functions $f_{h}$ are a linear map of
$\widetilde{E_{in}}\to\widetilde{E_{out}}$~\cite{MSSurveyAllFunctionsAndTypes}:
\begin{equation}
f_{h}\left(c\cdot\overrightarrow{E_{in}}+\sum_{\forall k}c_{k}\cdot\widetilde{E_{in}^{k}}\right)=c\cdot f_{h}\left(\overrightarrow{E_{in}}\right)+\sum_{\forall k}c_{k}\cdot f_{h}\left(\widetilde{E_{in}^{k}}\right)\label{eq:linerarityFIELD}
\end{equation}
where $k$ is any index, and $c,\,c_{k}\in\mathcal{R}$.
\end{rem}
The linearity property, in conjunction with the symmetry property
will be promptly employed to reform the input/output format of $f_{h}$,
without loss of generality.

\subsection{Specialized Modeling of Function Inputs/Outputs\label{subsec:modelIO}}

\begin{figure}[t]
\begin{centering}
\includegraphics[viewport=0bp 0bp 600bp 460bp,clip,width=0.8\columnwidth]{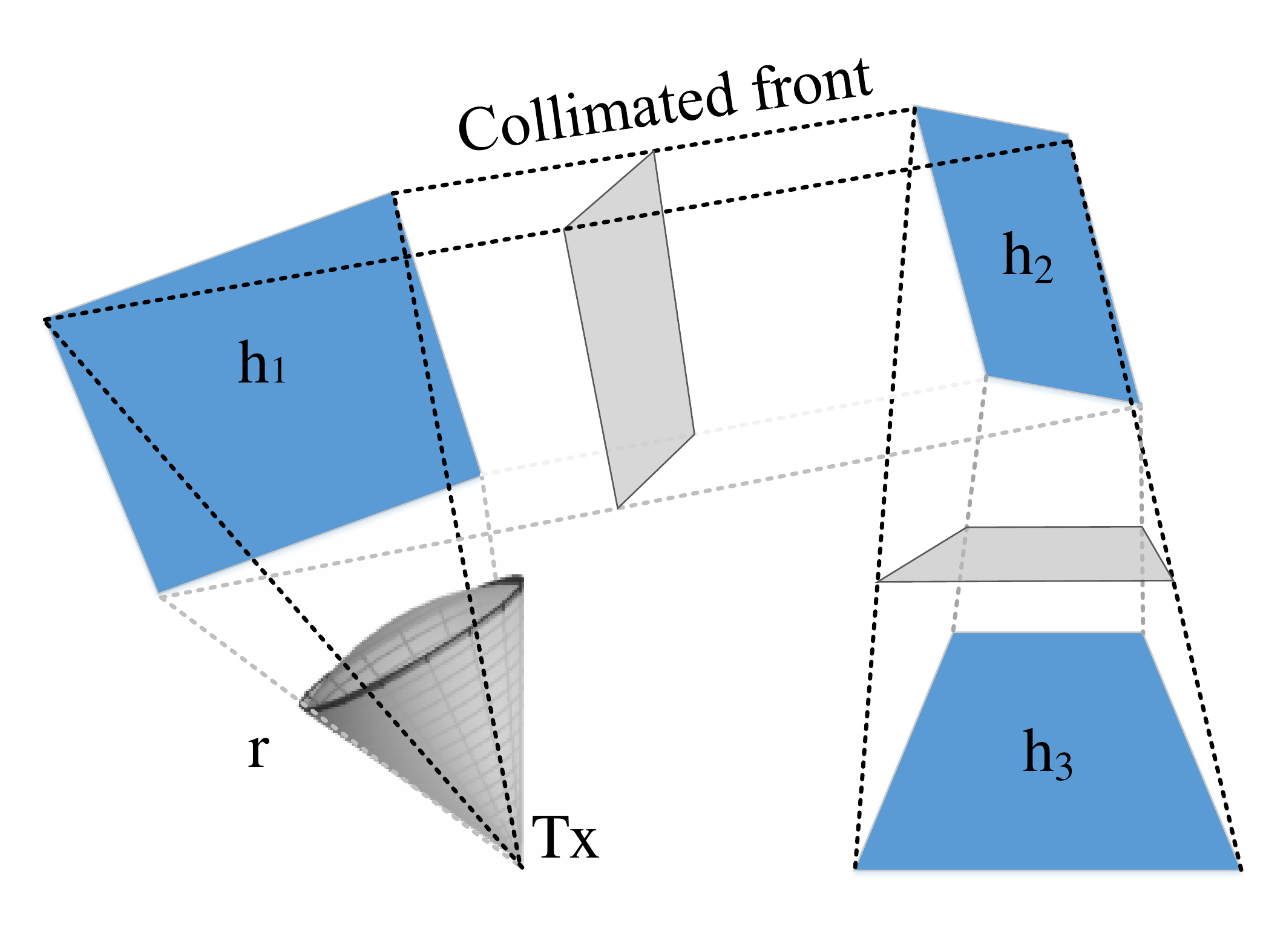}
\par\end{centering}
\caption{\label{fig3}Illustration of the propagation wavefront between users
and tiles. }
\end{figure}
In communication scenarios, considering function input/outputs at
the level of EM field may not be practical. Instead, considering the
signal source location and characteristics that yields the $\widetilde{E_{in}}$
can be more useful. To this end, we define the following input formats:

\textbf{Planar wave.} This case corresponds to a wave with a planar
or almost planar wavefront, as shown in Fig.~\ref{fig3}.

Planar waves can approximate:
\begin{itemize}
\item waves impinging on a tile from a distant antenna at its far region,
\item waves that have been collimated at a preceding tile ($e.g.,h_{1}$
in Fig.~\ref{fig3}), by applying the corresponding EM function.
\end{itemize}
Focusing on the second case, we will treat the collimation output
as the source of the planar wave. Employing the coordinate system
of the tile receiving this wave ($h_{2}$ in Fig.~\ref{fig3}) and
due to the planarity assumption, a single-frequency ($\omega$) wave
of this class can be simply described by:
\begin{itemize}
\item its direction, $\overline{D}:\left\{ r=\emptyset,\,\theta,\,\phi\right\} $
(using the $\emptyset$ notation to denote irrelevance from the $r$-dimension).
\item the total carried power, $\mathbf{P}$, that impinges upon the surface
of the tile (the summation of the Poynting vector norm over any bounded
wavefront),
\item the normalized Jones vector, $\overrightarrow{\mathbf{J}}$ ~\cite{jonesVector},
describing the wave polarization at tile~$h_{2}$.
\item the wavefront phase $\Phi$ at tile~$h_{2}$.
\end{itemize}
Since the field $\widetilde{E_{in}}$ can be reconstructed from the
aforementioned quantities, we proceed to replace it with the input
parameter set for plane waves, $\widetilde{\mathbb{I}_{PW}}$:
\begin{equation}
\widetilde{E_{in}}\mapsto\widetilde{\mathbb{I}_{PW}}:\,\left\langle \omega,\,\overline{D},\,\mathbf{P},\,\overrightarrow{\mathbf{J}},\,\Phi\right\rangle \label{eq:EinReplace}
\end{equation}
Likewise, a planar output field $\widetilde{E_{out}}$ \emph{defined
over the surface of $h_{2}$} can be alternatively expressed by the
output parameter set for plane waves, $\widetilde{\mathbb{O}_{PW}}$:
\begin{equation}
\widetilde{E_{out}}\mapsto\widetilde{\mathbb{O}_{PW}}:\,\left\langle \omega',\,\overline{D}',\,\mathbf{P}',\,\overrightarrow{\mathbf{J}}',\,\Phi'\right\rangle \label{eq:EoutReplace}
\end{equation}
where the $\left('\right)$ notation implies generally different values
than relation (\ref{eq:EinReplace}).

\textbf{Focal wave.} This case represents any generally non-collimated
radiation from a mobile device (cf. \noun{Tx} in Fig.~\ref{fig3}),
with its energy dissipating over an ever-growing sphere. In this case,
the EM field impinging upon a tile also depends on the characteristics
of the antenna device (radiation pattern and orientation). Assuming
that this information is known and constant, and using a similar approach
as above an input field at a tile, generated by such a source can
be replaced as:
\begin{equation}
\widetilde{E_{in}}\mapsto\widetilde{\mathbb{I}_{FW}}:\,\left\langle \omega,\,\overline{D},\,\mathbf{P},\,\overrightarrow{\mathbf{J}},\,\Phi\right\rangle \label{eq:EinReplaceFW}
\end{equation}
where $\overline{D}:\left\{ r,\,\theta,\,\phi\right\} $, and $\overrightarrow{\mathbf{J}},\,\Phi$
potentially vary over the tile surface. (I.e., $\overrightarrow{\mathbf{J}}\left(r,\,\theta\right),\,\Phi\left(r,\,\theta\right)$).
In the case where the focal wave is created (output) by a tile with
the application of a proper function, the output field is similarly
expressed as:
\begin{equation}
\widetilde{E_{out}}\mapsto\widetilde{\mathbb{O}_{FW}}:\,\left\langle \omega',\,\overline{D}',\,\mathbf{P}',\,\overrightarrow{\mathbf{J}}',\,\Phi'\right\rangle \label{eq:EoutReplaceFW}
\end{equation}

\begin{rem}
The notations (\ref{eq:EinReplace}),~(\ref{eq:EoutReplace}),~(\ref{eq:EinReplaceFW})
and (\ref{eq:EoutReplaceFW}) are compatible with the Symmetry property
(Remark~\ref{rem:F_symmetry-}), since $\mathbb{O}_{PW}$ and $\mathbb{O}_{FW}$
are also valid as inputs of any function $f_{h}$, while $\mathbb{I}_{PW}$
and $\mathbb{I}_{FW}$ are also valid as outputs of any function $f_{h}$
at tile~$h$.
\end{rem}
Finally, the linear map property of Remark~\ref{rem:F_linearity},
relation (\ref{eq:linerarityFIELD}) is rewritten as:
\begin{equation}
f_{h}\left(\left\{ \mathbb{I}_{1},\,\mathbb{I}_{2},\,\mathbb{I}_{3},\dots\right\} \right)=\left\{ f_{h}\left(\mathbb{I}_{1}\right),\,f_{h}\left(\mathbb{I}_{2}\right),\,f_{h}\left(\mathbb{I}_{3}\right),\dots\right\} \label{eq:linerarityFIELDrewrite}
\end{equation}
meaning that when multiple inputs are passed to a tile function, the
total output is produced by applying the function to each input separately.

\subsection{Modeling Core HyperSurface Functions\label{subsec:Modeling-Core-Functions}}

We proceed to study specialized HyperSurface functions, which are
of practical value to the studied programmable wireless environments.
Pure functions that can act as building blocks will be studied first,
followed by a model for combining them into more complex ones.

\textbf{Absorb} $f_{h}^{\text{\textsc{Abs}}}$. Plane wave absorption
has constituted one the most prominent showcases of metasurfaces~\cite{MSSurveyAllFunctionsAndTypes}.
In the studied programmable environments, absorbing unwanted reflections
is important for interference minimization, as well as enforcing determinism
over EM propagation. The ideal absorption function for an intended
plane wave input is expressed as:
\begin{equation}
f_{h}^{\text{\textsc{Abs}}}\left(\mathbb{I}_{PW}\right)\to\emptyset\label{eq:FabsInt}
\end{equation}
where the empty set denotes no output wave. Full absorption is attained
by matching the surface impedance of the tile to the incoming wave.
For a planar input, this means that the surface impedance is constant
across the tile, resulting into zero phase gradient and normal (specular)
reflective behavior. This remark facilitates the modeling of unintended
inputs as follows:
\begin{multline}
f_{h}^{\text{\textsc{Abs}}}\left(\widetilde{\mathbb{I}_{PW}}:\,\left\langle \omega,\,\overline{D},\,\mathbf{P},\,\overrightarrow{\mathbf{J}},\,\Phi\right\rangle ,\,\mathbb{I}_{PW}\right)\to\\
\left\langle \omega,\,\text{\textsc{Spec}}\left(\overline{D}\right),\,g\cdot\mathbf{P},\,\overrightarrow{\mathbf{J}},\,\Phi\right\rangle \label{eq:FabsUnint}
\end{multline}
where $g\left(\mathbb{I}_{PW},\,\widetilde{\mathbb{I}_{PW}}\right)\in\mathcal{R}<1$
is a metric of similarity between $\mathbb{I}_{PW}$ and $\widetilde{\mathbb{I}_{PW}}$,
defined by the physical structure of the HyperSurface and incorporating
any constant material losses. The specular reflection of $\overline{D}$
is calculated as:
\begin{equation}
\text{\textsc{Spec}}\left(\overline{D}\right)=\overline{D}-2\left(\overline{D}\cdot\vec{n}\right)\vec{n}\label{eq:reflectionGeometric}
\end{equation}
where $\overline{D}$ is directed and $\vec{n}$ is the unit normal
of the tile surface.

It is noted that relation (\ref{eq:FabsUnint}) can be employed to
achieve an intended: i) partial attenuation of a wave, and ii)~frequency
selective absorption (filtering)~\cite{MSSurveyAllFunctionsAndTypes}.

\textbf{Steer} $f_{h}^{\text{\textsc{Str}}}$. Steering plain waves
from an incoming direction to another is achieved by enforcing a gradient
surface impedance that corresponds to the required reflection index~\cite{MSSurveyAllFunctionsAndTypes}.
This physical phenomenon can be described by considering a virtual
surface normal, $\vec{n}'$, supplied as an input parameter of steering,
that corresponds to a required reflection direction via relation~(\ref{eq:reflectionGeometric}).
We proceed to define $f_{h}^{\text{\textsc{Str}}}$ as follows:
\begin{multline}
f_{h}^{\text{\textsc{Str}}}\left(\widetilde{\mathbb{I}_{PW}}:\,\left\langle \omega,\,\overline{D},\,\mathbf{P},\,\overrightarrow{\mathbf{J}},\,\Phi\right\rangle ,\,\vec{n}',\,\mathbb{I}_{PW}\right)\to\\
\left\langle \omega,\,\overline{D}-2\left(\overline{D}\cdot\vec{n}'\right)\vec{n}',\,g\cdot\mathbf{P},\,\overrightarrow{\mathbf{J}},\,\Phi\right\rangle \label{eq:FstrUnint}
\end{multline}
where $g\left(\mathbb{I}_{PW},\,\widetilde{\mathbb{I}_{PW}}\right)\in\mathcal{R}<1$
in defined as in relation~(\ref{eq:FabsUnint}), noting that its
specific expression generally differs from that of relation~(\ref{eq:FabsUnint}).
We remark that the variant surface normal approach allows for a uniform
expression~(\ref{eq:FstrUnint}), covering both the intended and
unintended inputs.

\textbf{Collimate} $f_{h}^{\text{\textsc{Col}}}$. As noted in Section~\ref{sec:Related-work},
collimation is the action of transforming a non-planar input wave
to a planar output~\cite{MSSurveyAllFunctionsAndTypes}. In essence,
the HyperSurface is configured for a virtual surface normal that varies
across the tile surface, matching the local direction of arrival of
the input. In the present scope, collimation will be studied for focal
wave input as follows:
\begin{equation}
f_{h}^{\text{\textsc{Col}}}\left(\mathbb{I}_{FW},\,\overline{D}'\right)\to\mathbb{O}_{PW}:\,\left\langle \omega,\,\overline{D}',\,g\cdot\mathbf{P},\,\overrightarrow{\mathbf{J}},\,\Phi\right\rangle \label{eq:CollIntend}
\end{equation}
where the focal wave characteristics and the intended reflection direction
are provided as inputs. Relation~(\ref{eq:CollIntend}) refers to
intended inputs and outputs. Obtaining the output to an unintended
input can be based on the surface normal across the tile. Each sub-area
over which the surface normal can be considered constant interacts
with a part of the wavefront that can be considered planar, resulting
into a reflection calculated via relation~(\ref{eq:reflectionGeometric}).
Thus, the outcome to unintended input is a set of planar outputs,
each with its own power, polarity and phase:
\begin{equation}
f_{h}^{\text{\textsc{Col}}}\left(\widetilde{\mathbb{I}_{FW}},\,\mathbb{I}_{FW},\,\overline{D}'\right)\to\left\{ \mathbb{O}_{PW}\right\} \label{eq:CollUnintnd}
\end{equation}
In the context of the studied programmable environments, collimation
is intended to be used mainly at the first and last hop of the propagation
from one device to another~\cite{Liaskos2019ADHOC}. The first application
(at a tile) transforms the waves emitted from a device to a planar
form (see Fig.~\ref{fig3}). Tile-to-tile propagation is then performed
for planar inputs. At the final tile before reaching the receiver,
the planar wave is focused to the intended spot. This focusing is
essentially collimation in the reverse, where a planar wave is converted
to focal output. The tile configuration for focusing remains the same
as in collimation, due to the symmetry in Remark~\ref{rem:F_symmetry-}.

\textbf{Polarize} $m_{h}^{\text{\textsc{Pol}}}$. The physics of polarization
control can be described qualitatively by assuming equivalent circuits
of meta-atoms~\cite{MSSurveyAllFunctionsAndTypes}. In essence, each
meta-atom can be viewed as a circuit with a input and output antennas,
as well as cross-couplings among meta-atoms. A wave enters via the
input antennas, undergoes some alteration via the circuit and exits
via output antennas, subject to the connections performed by tuning
the active elements. Polarization is thus a shift in the Jones vector,
attained by the appropriate choice of output antennas. Pure polarization
control does not affect other wave parameters. As such, we define
the polarization control not as a stand-alone function, but rather
as a modifier applied to the output of a preceding function:
\begin{multline}
m_{h}^{\text{\textsc{Pol}}}\left(\mathbb{O}_{PW}:\left\langle \omega,\,\overline{D},\,\mathbf{P},\,\overrightarrow{\mathbf{J}},\,\Phi\right\rangle ,\,\Delta\overrightarrow{\mathbf{J}}\right)\to\\
\mathbb{O}'_{PW}:\left\langle \omega,\,\overline{D},\,\mathbf{P},\,\overrightarrow{\mathbf{J}}+\Delta\overrightarrow{\mathbf{J}},\,\Phi\right\rangle \label{eq:PollIntended}
\end{multline}
For unintended outputs, the intended shift $\Delta\overrightarrow{\mathbf{J}}$
is not necessarily attained, which can be expressed as:
\begin{multline}
m_{h}^{\text{\textsc{Pol}}}\left(\widetilde{\mathbb{O}_{PW}},\,\mathbb{O}_{PW},\,\Delta\overrightarrow{\mathbf{J}}\right)\to\\
\widetilde{\mathbb{O}'_{PW}}:\left\langle \omega,\,\overline{D},\,\mathbf{P},\,\overrightarrow{\mathbf{J}}+\Delta\overrightarrow{\mathbf{J}}',\,\Phi\right\rangle \label{eq:PollUnIntended}
\end{multline}
where $\,\Delta\overrightarrow{\mathbf{J}}'\ne\Delta\overrightarrow{\mathbf{J}}$
in general. Notice that power loss concerns are delegated to the preceding
pure function.

\textbf{Phase Alteration} $m_{h}^{\text{\textsc{Pha}}}$. Phase alteration
follows the same principle as the polarization. Qualitatively, the
modification of the wave phase is accomplished within the equivalent
circuit, via an inductive or capacitative element~\cite{MSSurveyAllFunctionsAndTypes}.
Once again, this functionality is defined as a modifier:
\begin{multline}
m_{h}^{\text{\textsc{Pha}}}\left(\mathbb{O}_{PW}:\left\langle \omega,\,\overline{D},\,\mathbf{P},\,\overrightarrow{\mathbf{J}},\,\Phi\right\rangle ,\,\Delta\Phi\right)\to\\
\mathbb{O}'_{PW}:\left\langle \omega,\,\overline{D},\,\mathbf{P},\,\overrightarrow{\mathbf{J}},\,\Phi+\Delta\Phi\right\rangle \label{eq:PhaseIntended}
\end{multline}
For unintended outputs, the definition is altered similarly to relation~(\ref{eq:PollUnIntended}).

\subsubsection*{Combination model}

The pure functions studied above can be combined to describe a more
complex functionality. Complex functions may be an operational requirement
(e.g., steer and polarize at the same time), or be imposed by physical
imperfections of the metasurface (e.g., being unable to steer without
altering the polarization). We proceed to present a common model to
describe both cases.

\textbf{Surface division}. This combination approach assigns different
functions to different sub-areas of the same tile~\cite{CUIcoding2018}.
The principle of operation is shown in Fig.~\ref{fig4} (left inset),
where an impinging plane undergoes splitting into 3 separate directions
and partial attenuation. The power distribution of any output waves
depends on the area allocated to each sub-function. For instance,
an ideal N-way split of an input $\mathbb{I}_{PW}$ towards directions
with custom normals $\vec{n}_{i}',\,i=1\ldots N$ can be expressed
as:
\begin{figure}[t]
\begin{centering}
\includegraphics[viewport=0bp 0bp 730bp 240bp,clip,width=1\columnwidth]{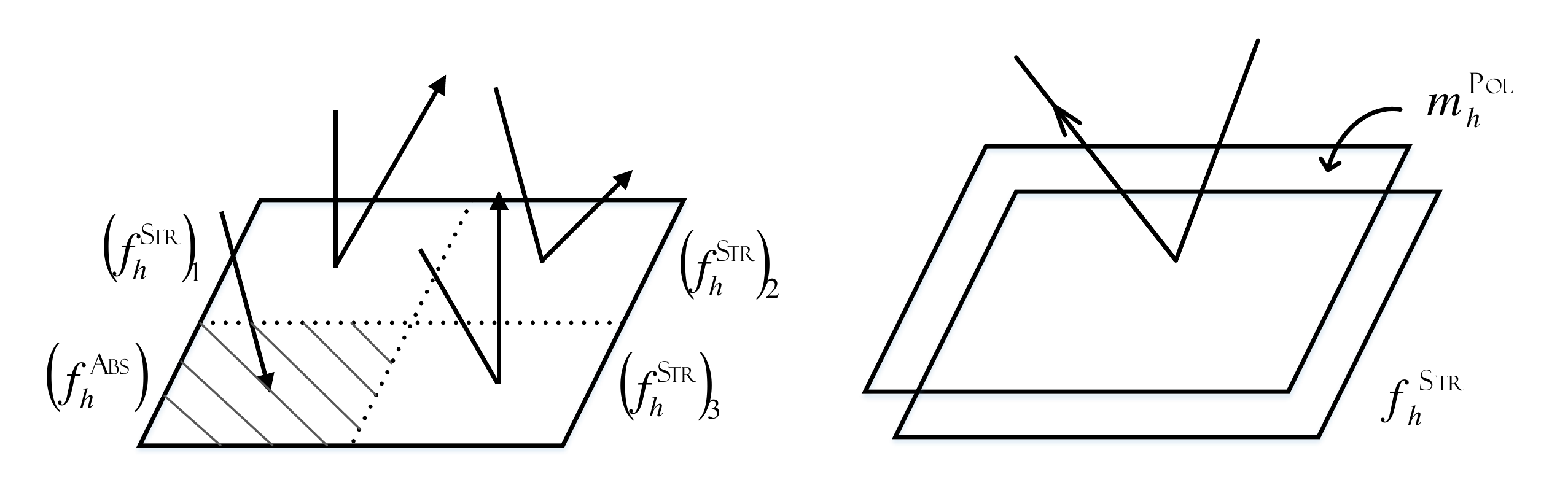}
\par\end{centering}
\caption{\label{fig4}Function combination models: Surface Division (left)
and Meta-atom Merge (right).}
\end{figure}
\begin{multline}
f_{h}^{\text{\textsc{Spl}}}\left(\mathbb{I}_{PW},\left\{ \vec{n}_{i}'\right\} \right)\to\left\{ \mathbb{O}_{PW}\right\} ,\\
\left\{ \mathbb{O}_{PW}\right\} :\,\left\{ {\scriptstyle \left\langle \omega,\,\overline{D}-2\left(\overline{D}\cdot\vec{n_{i}}'\right)\vec{n_{i}}',\,\frac{1}{N}\cdot\mathbf{P},\,\overrightarrow{\mathbf{J}},\,\Phi\right\rangle }\text{\ensuremath{\vert\,i=1\ldots N}}\right\} \label{eq:Fspl}
\end{multline}
where $\left(\nicefrac{\mathbf{P}}{N}\right)$ is the power of each
output.

\textbf{Meta-atom merge}. A common practice in metasurface analysis
is to merge meta-atoms to create more complex basic structures, called
supercells~\cite{Holloway.2012}. Essentially, using the equivalent
circuit paradigm, merging meta-atom creates circuits with more degrees
of tunability. This enables the combination of functions and modifiers,
e.g., for steering and polarizing or steering and phase altering at
the same time, over the same surface (cf.~Fig.~\ref{fig4}, right
inset), denoted as:
\begin{equation}
m_{h}^{\text{\textsc{Pol}}}\left(f_{h}^{\text{\textsc{Str}}}\right),\,m_{h}^{\text{\textsc{Pha}}}\left(f_{h}^{\text{\textsc{Str}}}\right),\,\text{or\,}m_{h}^{\text{\textsc{Pol}}}\left(m_{h}^{\text{\textsc{Pha}}}\left(f_{h}^{\text{\textsc{Str}}}\right)\right)
\end{equation}
Additionally, it is possible to apply a meta-atom merge-derived function
only to a sub-area of a tile, thus combining it with surface division.

A summary of combinable functions and combination approaches is given
in Table~\ref{tab:combFunctions}. Notably: i)~the combination patterns
are symmetric, ii)~combining collimation with any other non-modifier
function is possible but potential unintended, as described in the
context of relation~(\ref{eq:CollUnintnd}), iii)~combining a modifier
with absorption makes sense only when the absorption is partial (i.e.,
there is an output to apply the modifier upon), and iv)~combining
several modifiers of the same type to the same tile is the same as
applying it once with the total modification. As such, these combinations
are trivial. \emph{Finally, it is noted that any number of functions
(i.e., more than two) can be merged via the surface division model.}

\begin{table}[t]
\begin{centering}
\caption{\textsc{\label{tab:combFunctions}Combinable tile functions and methods.}}
\begin{tabular}{|c|c|c|c|c|c|c|}
\cline{3-7}
\multicolumn{1}{c}{} &  & \multicolumn{5}{c|}{\noun{Merge with}}\tabularnewline
\cline{3-7}
\multicolumn{1}{c}{} & $\phantom{\int^{\int}}$ & $f_{h}^{\text{\textsc{Abs}}}$ & $f_{h}^{\text{\textsc{Str}}}$ & $f_{h}^{\text{\textsc{Col}}}$ & $m_{h}^{\text{\textsc{Pol}}}$ & $m_{h}^{\text{\textsc{Pha}}}$\tabularnewline
\hline
\multirow{10}{*}{\begin{turn}{90}
\noun{Function}
\end{turn}} & \multirow{2}{*}{$f_{h}^{\text{\textsc{Abs}}}$} & \multirow{2}{*}{SD} & \multirow{2}{*}{SD} & \multirow{2}{*}{$\utilde{\text{SD}}$} & \multirow{2}{*}{MM$^{*}$} & \multirow{2}{*}{MM$^{*}$}\tabularnewline
 &  &  &  &  &  & \tabularnewline
\cline{2-7}
 & \multirow{2}{*}{$f_{h}^{\text{\textsc{Str}}}$} & \multirow{2}{*}{SD} & \multirow{2}{*}{SD} & \multirow{2}{*}{$\utilde{\text{SD}}$} & \multirow{2}{*}{MM} & \multirow{2}{*}{MM}\tabularnewline
 &  &  &  &  &  & \tabularnewline
\cline{2-7}
 & \multirow{2}{*}{$f_{h}^{\text{\textsc{Col}}}$} & \multirow{2}{*}{$\utilde{\text{SD}}$} & \multirow{2}{*}{$\utilde{\text{SD}}$} & \multirow{2}{*}{$\utilde{\text{SD}}$} & \multirow{2}{*}{MM} & \multirow{2}{*}{MM}\tabularnewline
 &  &  &  &  &  & \tabularnewline
\cline{2-7}
 & \multirow{2}{*}{$m_{h}^{\text{\textsc{Pol}}}$} & \multirow{2}{*}{MM$^{*}$} & \multirow{2}{*}{MM} & \multirow{2}{*}{MM} & \multirow{2}{*}{-} & \multirow{2}{*}{MM}\tabularnewline
 &  &  &  &  &  & \tabularnewline
\cline{2-7}
 & \multirow{2}{*}{$m_{h}^{\text{\textsc{Pha}}}$} & \multirow{2}{*}{MM$^{*}$} & \multirow{2}{*}{MM} & \multirow{2}{*}{MM} & \multirow{2}{*}{MM} & \multirow{2}{*}{-}\tabularnewline
 &  &  &  &  &  & \tabularnewline
\hline
\end{tabular}
\par\end{centering}
\begin{centering}
\textbf{SD:} Surface Division, \textbf{MM:} Meta-atom Merge,
\par\end{centering}
\begin{centering}
$\sim$: Possible but potentially unintended.
\par\end{centering}
\centering{}{*}: Defined only when $f_{h}^{ABS}$ produces output
(partial absorb).
\end{table}
It is worth noting that merging functions is not without impact on
the efficiency of the HyperSurface. Any metasurface requires a minimum
amount of meta-atoms to yield a consistent behavior (i.e., with near-unitary
efficiency)~\cite{CUIcoding2018}. Both surface division and meta-atom
merge naturally limit the meta-atom numbers available for a given
functionality. As such, combining functions will generally amplify
discretization, boundary and other errors~\cite{RadarDiscrete}.
This can lead to reduced efficiency, which can be expressed as considerably
attenuated intended outputs, as well as the appearance of unintended,
parasitic outputs, $\mathbb{O_{P}}$, even for intended inputs:
\begin{equation}
f_{h}\left(\mathbb{I},\widetilde{\mathbb{I}}\right)\to f_{h}\left(\mathbb{I}\right)+f_{h}\left(\widetilde{\mathbb{I}}\right)+\mathbb{O_{P}}\label{eq:profile}
\end{equation}
where $f_{h}\left(\mathbb{I}\right)$ is the intended output and $f_{h}\left(\widetilde{\mathbb{I}}\right)$
is the well-defined output to unintended input. Thus:
\begin{rem}
\label{rem:DONOTREUSE}Combining functionalities over a single tile
generally reduces the efficiency of the overall function.
\end{rem}
The effects of Remark~\ref{rem:DONOTREUSE} can be quantified only
per specific, physical HyperSurface design. Nonetheless, we will employ
this remark in ensuing Sections and setup a policy of minimizing function
combinations in programmable wireless environments.
\begin{rem}
\label{rem:PROFILE}The generic relation~(\ref{eq:profile}), in
conjunction with the preceding modeling defines the information that
a HyperSurface manufacturer should measure and provide, to facilitate
the use in programmable environments. This information, collectively
referred to as \emph{EM profile} contains the following information:
\end{rem}
\begin{itemize}
\item The supported function types and allowed combinations, as a subset
of the entries of Table~\ref{tab:combFunctions}.
\item The intended and parasitic outputs to intended inputs.
\item The unintended and parasitic outputs to unintended inputs.
\end{itemize}
The EM profile can be obtained, e.g., by measuring the scattering
pattern in controlled conditions, for the complete array of supported
functions and intended inputs. Nonetheless, an exhaustive measurement
for any unintended input may be prohibitive. In this case, the profile
may provide a calculation model for outputs, such as the $g\left(\mathbb{I}_{PW},\,\widetilde{\mathbb{I}_{PW}}\right)$
metric employed in Section~\ref{subsec:Modeling-Core-Functions}.

For the remainder of this study, we will consider the EM profile as
a given, provided by the tile manufacturer.

\subsection{A Graph Model for Simulating and Optimizing Programmable Wireless
Environments}

Propagation within a 3D space comprises Line-of-sight (LOS) and Non-LOS
(NLOS) components. Naturally, PWEs control the NLOS component only,
without affecting the LOS~\cite{Liaskos:2018:UAS:3289258.3192336}.
Therefore, the following model will focus on the NLOS case. It is
noted that a workaround for total control over NLOS and LOS with PWEs
is discussed in Section~\ref{sec:Discussion-and-Future}.

Consider two tiles, $h$ and $h'$, in a 3D space. We will consider
these tiles as \emph{connectable} if there exist any input $\widetilde{\mathbb{I}}$
and functions $f_{h}$,~$f_{h'}$, such that:
\begin{equation}
f_{h'}\left(f_{h}\left(\mathbb{\widetilde{\mathbb{I}}}\right)\right)\to\emptyset\,\iff f_{h'}:f_{h'}^{ABS},\,\forall f_{h}\left(\mathbb{\widetilde{\mathbb{I}}}\right)\ne\emptyset\label{eq:InterTile}
\end{equation}
In other words, inter-tile connectivity means that one tile can redirect
impinging EM energy to another. It is implied that the redirected
power in significant, i.e., it surpasses a practical threshold defined
by the application scenario. Thus, connectability may be precluded
due to physical obstacles between tiles, or by the lack of supported
tile functions to redirect significant energy to one another.

Additionally, we consider a set of user devices, $\mathcal{U}$, in
the same space. User $u\in\mathcal{U}$ will be considered \emph{connected}
to tile $h$ if there exists any LOS input $\widetilde{\mathbb{I}}$
and a function $f_{h}$ such that:
\begin{equation}
f_{h}\left(\mathbb{\widetilde{\mathbb{I}}}\right)\to\emptyset\,\iff f_{h}:f_{h}^{ABS},\,\forall\mathbb{\widetilde{\mathbb{I}}}\ne\emptyset\label{eq:TileToUser}
\end{equation}
i.e., getting zero output from a tile connected to a user is only
possible if a full absorption functionality is employed.

Based on relations (\ref{eq:InterTile}) and \eqref{eq:TileToUser},
we define the notion of:
\begin{itemize}
\item the set of inter-tile links, $\mathcal{L}_{h}$, where each contained
link $l_{h,h'}\in\mathcal{L}_{h}$ denotes tile \emph{connectivity
potential} by relation~\eqref{eq:InterTile}, and
\item the set of user-tile links, $\mathcal{L}_{u}$, where each contained
link $l_{u,h}\in\mathcal{L}_{u}$ denotes \emph{connection} by relation~\eqref{eq:TileToUser}.
\end{itemize}
Following the symmetry Remark~\ref{rem:F_symmetry-}, all links in
$\mathcal{L}_{h}$ and $\mathcal{L}_{u}$ are bidirectional and symmetric.
Additionally all links will be considered to have an associated label,
$\text{\textsc{delay}}\left(l\right)$, defined as the wave propagation
delay, inclusive of any delay within the receiving end. Finally, last-tile-\emph{to}-receiving-user
links may be labeled as $\text{\textsc{Tx}}\left(l_{u,h}\right)\in\mathcal{U}$
, to designate the transmitting user that may employ them. Likewise,
transmitting-user-\emph{to}-first-tile links may be labeled as $\text{\textsc{Rx}}\left(l_{u,h}\right)\in\mathcal{U}$
, to designate the receiving user that must be reached via them. The
labeling is intended to capture the Multiple-Input-Multiple-Output
(MIMO) potential of the user devices at a high level.

Based on the above definitions, we proceed to define the graph $\mathcal{G}\left\langle \left\{ \mathcal{H},\,\mathcal{U}\right\} ,\,\left\{ \mathcal{L}_{h},\,\mathcal{L}_{u}\right\} \right\rangle $,
as well as subgraphs of the form $\mathcal{G}\left(\left\{ f_{h}\right\} \right)$.
The latter represent environments that have been configured by applying
specific functions to tiles, thus instantiating some of the allowed
links $\mathcal{L}_{h}$. We consider an EM flow, $\left\langle u_{\text{\textsc{tx}}},u_{\text{\textsc{rx}}}\right\rangle $,
from a transmitter $u_{\text{\textsc{tx}}}$ to a receiver $u_{\text{\textsc{rx}}}$,
following a path via tiles within $\mathcal{G}\left(\left\{ f_{h}\right\} \right)$,
defined as an ordered selection of links without repetitions:
\begin{equation}
\overrightarrow{p}_{\left\langle u_{\text{\textsc{tx}}}u_{\text{\textsc{rx}}}\right\rangle }=\left\{ l_{u_{\text{\textsc{tx}}}h_{1}},\,l_{h_{1}h_{2}},\,l_{h_{2}h_{3}},\,\ldots,l_{h_{N}u_{\text{\textsc{rx}}}}\right\} .\label{eq:pathdef}
\end{equation}

\begin{rem}
The above formulation with non-repeating links in paths is also posed
in compliance with Remark~\ref{rem:DONOTREUSE}, which dictates the
avoidance of function combinations. To understand this claim, consider
a counter-example with a repetitive path. This path must have the
form $\left\{ \ldots l_{h_{k}h_{i}},\,l_{h_{i}h_{i+1}},\,\dots l_{h_{l}h_{i}},\,l_{h_{i}h_{i+1}},\ldots\right\} $
where $k\ne l$. Due to the symmetry Remark~\ref{rem:F_symmetry-},
a reverse input via the repeated link $l_{h_{i}h_{i+1}}$ must exit
from both $l_{h_{k}h_{i}}$ and $l_{h_{l}h_{i}}$ . This requires
a splitting function at tile $h_{i}$, which is a combined function,
as described in the context of relation~(\ref{eq:Fspl}).
\end{rem}
Any input $\mathbb{\widetilde{\mathbb{I}}}$ at tile $h_{1}$ is transformed
over the path as:
\begin{equation}
\widetilde{\mathbb{O}}\left(\widetilde{\mathbb{I}},\,\overrightarrow{p}_{\left\langle u_{\text{\textsc{tx}}}u_{\text{\textsc{rx}}}\right\rangle }\right)\gets f_{h_{N}}\circ\ldots f_{h_{3}}\circ f_{h_{2}}\circ f_{h_{1}}\circ\widetilde{\mathbb{I}}\label{eq:pathO}
\end{equation}
where $f_{h_{j}}\circ f_{h_{i}}$ implies passing only the output
of $f_{h_{i}}$ that impinges tile $h_{j}$ to $f_{h_{j}}$, and ignoring
any other outputs of $f_{h_{i}}$. Without loss of generality, relation~(\ref{eq:pathO})
omits the propagation over the first and last link, which are subject
to antenna characteristics and standard (non-programmable) propagation.
The end-to-end delay of a path can be calculated in a trivial manner
as the sum of all link delays:
\begin{equation}
\text{\textsc{Delay}\ensuremath{\left(\overrightarrow{p}_{\left\langle u_{\text{\textsc{tx}}}u_{\text{\textsc{rx}}}\right\rangle }\right)}}=\sum_{\forall l\in\overrightarrow{p}}\textsc{Delay}\left(l\right)\label{eq:delayPath}
\end{equation}
We proceed to consider all paths that reach the receiver $u_{\text{\textsc{rx}}}$,
from any transmitter $(*)$ (noting the possible multiplicity of paths
for the same $\left\langle u_{\text{\textsc{tx}}},u_{\text{\textsc{rx}}}\right\rangle $
pair). The total received output is then:
\begin{equation}
\widetilde{\mathbb{O}}_{u_{\text{\textsc{rx}}}}^{\text{\textsc{total}}}=\sum_{\forall\overrightarrow{p}:\overrightarrow{p}\left\langle *,u_{\text{\textsc{rx}}}\right\rangle }\widetilde{\mathbb{O}}\left(\widetilde{\mathbb{I}},\,\overrightarrow{p}\right)\label{eq:totalOut}
\end{equation}
Relations (\ref{eq:delayPath}), (\ref{eq:totalOut}) contain all
the required information for deriving the communication quality of
a pair $\left\langle u_{\text{\textsc{tx}}},u_{\text{\textsc{rx}}}\right\rangle $.
Moreover, due to the symmetry Remark~\ref{rem:F_symmetry-}, the
derivations of relations (\ref{eq:delayPath}), (\ref{eq:totalOut})
are identical for the reverse path, $\left\langle u_{\text{\textsc{rx}}},u_{\text{\textsc{tx}}}\right\rangle $.
Finally, the labeling $\text{\textsc{Tx}}\left(l_{h_{N}u_{\text{\textsc{rx}}}}\right)$
of the receiver's links can be employed to classify outputs as useful
or interfering based on the MIMO configuration.

\begin{algorithm}[t]
\begin{algorithmic}[1]

\Procedure{$\text{\texttt{\textsc{paths}}\ensuremath{{}_{\text{\textsc{rx}}}}}$:
NLOSProp}{$\mathcal{G}\left(\left\{ f_{h}\right\} \right),u_{\text{\textsc{rx}}}$,
$\left\{ \left\langle u_{\text{\textsc{tx}}},\left\{ \widetilde{\mathbb{I}}_{h},h:l_{u,h}\in\left\{ \mathcal{L}_{u}:u=u_{\text{\textsc{tx}}}\right\} \right\} \right\rangle \right\} $}

\State  $\texttt{rays}\gets\left\{ \right\} ,\,\text{\texttt{\textsc{paths}}\ensuremath{_{\text{\textsc{rx}}}}}\gets\left\{ \right\} $;

\State \textbf{for} $l_{u,h}\in\left\{ \mathcal{L}_{u}:u\in\left\{ u_{\text{\textsc{tx}}}\right\} \right\} $
~~\emph{//for all transmitters.}

\State  ~~$r\gets\left\langle \overrightarrow{p}_{r}=\left\{ l_{u,h}\right\} ,\widetilde{\mathbb{I}}_{r}=\widetilde{\mathbb{I}}_{h}\right\rangle $;

\State ~~$\texttt{rays}\gets\texttt{rays}+r$;

\State  \textbf{end for }

\State  \textbf{while} $\texttt{rays}\ne\emptyset$

\State ~~$r\gets\texttt{rays.pop\ensuremath{\left(\right)}}$;~~\emph{//remove
and return last entry.}

\State ~~\textbf{for} $o\in\widetilde{\mathbb{O}}\left(r:\widetilde{\mathbb{I}}_{r},\,r:\overrightarrow{p}_{r}\right)$~~\emph{//cf.
rel.~(\ref{eq:pathO}).}

\State ~~~~$h^{*}\gets\text{\textsc{tile\_reached}}\left(o\right)$;
~~\emph{//cf. rel.~(\ref{eq:InterTile}).}

\State ~~~~$u^{*}\gets\text{\textsc{user\_reached}}\left(o\right)$;\emph{~~//cf.
rel.~(\ref{eq:TileToUser}).}

\State ~~~~$h\gets\text{\textsc{last\_tile}}\left(r:\overrightarrow{p}_{r}\right)$;

\State ~~~~\textbf{if} $\left(u^{*}=u_{\text{\textsc{rx}}}\right)$

\State ~~~~\textbf{~~$r^{*}\gets\left\langle r:\overrightarrow{p}_{r}+l_{h,u^{*}},r:\widetilde{\mathbb{I}}_{r}\right\rangle ;$}

\State ~~~~~~$\texttt{\texttt{\textsc{paths}}\ensuremath{{}_{\text{\textsc{rx}}}}}\gets\texttt{\textsc{paths}}{}_{\text{\textsc{rx}}}+\left(r:\overrightarrow{p}_{r}+l_{h,u^{*}}\right)$;

\State ~~~~\textbf{else if} $\left(h^{*}\ne\emptyset\right)$

\State ~~~~\textbf{~~$r^{*}\gets\left\langle r:\overrightarrow{p}_{r}+l_{h,h^{*}},r:\widetilde{\mathbb{I}}_{r}\right\rangle ;$}

\State ~~~~~~$\texttt{rays}\gets\texttt{rays}+r^{*}$;

\State ~~~~\textbf{end if}

\State ~~\textbf{end for}

\State  \textbf{end while}

\EndProcedure

\end{algorithmic}

\caption{\label{alg:PropagationPWE}Simulated NLOS propagation model for PWEs.}
\end{algorithm}
The paths $\overrightarrow{p}_{\left\langle u_{\text{\textsc{tx}}}u_{\text{\textsc{rx}}}\right\rangle }$
are directly derived from the deployed tile functions and the corresponding
graph $\mathcal{G}\left(\left\{ f_{h}\right\} \right)$. The simulated
NLOS propagation procedure is modeled as \noun{NLOSprop} (Algorithm~\ref{alg:PropagationPWE}).
\textcolor{black}{The process receives as inputs: i) the configured
tiles, ii) a receiver, and iii) all transmitting users and their inputs
($\widetilde{\mathbb{I}}$ for each link connecting the transmitter
to a tile). The process produces the paths leading from any transmitter
to the receiver. Internally, a tupple denoted as }\textcolor{black}{\emph{ray}}\textcolor{black}{{}
is used for holding a path (cf. rel.~(\ref{eq:pathdef})) and its
input. A stack of $rays$ is initialized per transmitter link (lines
$3-6$). The model then continuously updates each ray, accounting
for tile interactions (line $9$). It is noted that each interaction
may yield more than one outputs. Once a path reaches the intended
user, it is added to the model outputs (lines $13-15$) and the corresponding
ray is discarded. If a tile is reached, a copy of the ray with an
updated path is added to the ray set for further processing (lines
$16-19$). The model accounts for attenuated rays, by using the connectivity
definitions~(\ref{eq:InterTile}),~(\ref{eq:TileToUser})\textendash weak
rays are treated as null outputs. Furthermore, notice that rays escaping
the consider space are discarded as expected (lines $10,11$ will
yield null outputs). Finally, notice that the model can also provide
the paths to a set of receivers, $\left\{ u_{\text{\textsc{rx}}}\right\} $,
by modifying the condition in line $13$ as $u^{*}\in\left\{ u_{\text{\textsc{rx}}}\right\} $.}
\begin{rem}
\label{rem:OptPWEsGeneric}Programmable wireless environments seek
to optimize the communication for any set $\left\langle \left\{ u_{\text{\textsc{rx}}}\right\} ,\left\{ u_{\text{\textsc{tx}}}\right\} \right\rangle $,
by deploying the corresponding, performance optimizing tile functions.
This can be generally formulated as:
\begin{multline}
\left\{ f_{h}\right\} ^{OPT}\gets\underset{\left\{ f_{h}\right\} }{\arg\text{opt}}(\text{\textsc{objective}}(\textsc{\ensuremath{{\scriptstyle \text{LOSProp}}}\ensuremath{\left(\left\langle \left\{  u_{\text{\textsc{rx}}}\right\}  ,\left\{  u_{\text{\textsc{tx}}}\right\}  \right\rangle \right)}}\\
+\text{\ensuremath{{\scriptstyle \textsc{NLOSProp}}}}\left(\mathcal{G}\left(\left\{ f_{h}\right\} \right),\left\langle \left\{ u_{\text{\textsc{rx}}}\right\} ,\left\{ u_{\text{\textsc{tx}}}\right\} \right\rangle \right)))\label{eq:OptGeneric}
\end{multline}
where \noun{objective} is a fitness function applied to the propagation
outcome, and the inputs $\widetilde{\mathbb{I}}$ of \noun{NLOSProp}
are omitted for clarity. The \noun{objective} may freely refer to:
\end{rem}
\begin{itemize}
\item $1$ transmitter to $1$ receiver (uni-cast), $\left\langle u_{\text{\textsc{tx}}},u_{\text{\textsc{rx}}}\right\rangle $,
\item 1 transmitter to many receivers (multi-cast or broad-cast), $\left\langle u_{\text{\textsc{tx}}},\left\{ u_{\text{\textsc{rx}}}\right\} \right\rangle $,
\item many transmitters to many receivers, $\left\langle \left\{ u_{\text{\textsc{tx}}}\right\} ,\left\{ u_{\text{\textsc{rx}}}\right\} \right\rangle $.
\end{itemize}
The latter two categories inherently incorporate resource sharing
policies among communicating pairs.

It is noted that the formulation of relation~(\ref{eq:OptGeneric})
is generic and, thus, also covers cases that are not compliant to
Remark~\ref{rem:DONOTREUSE} about tile functionality reuse.

\subsection{Modeling connectivity objectives\label{subsec:Modeling-connectivity-objectives}}

We proceed to study specific objectives for core aspects of wireless
system performance. We focus on objectives pertaining to a single
receiver, since resource sharing policies are subjective. It is noted,
however, that Section~\ref{sec:A-K-paths-Approach} case-studies
the integration of objectives and policies. Additionally, all \noun{LOS}
components will be omitted for clarity, since they are either not
affected by PWEs or be manipulated for control in the same manner
as NLOS, as described later in Section~\ref{sec:Discussion-and-Future}.

We proceed to define a path-centric formulation for various core objectives.
In other words, we employ the correspondence $\left\{ f_{h}\right\} \leftrightarrow\overrightarrow{p}$
between a set of deployed functions $f_{h}$ and formed paths $\overrightarrow{p}$
interconnecting the various users within the PWE. The practicality
of this choice will be explained in the remainder of the current subsection
and in Section~\ref{sec:A-K-paths-Approach}.

\subsubsection*{Power transfer maximization.}

We study the objective of maximizing the total power received by a
specific user, emitted by a group of transmitters. This objective
is practical for wireless power transfer~\cite{Liaskos:2018:UAS:3289258.3192336},
as well as for formulating subsequent objectives. Using a path-centric
approach, relation~(\ref{eq:OptGeneric}) can be rewritten as:
\begin{equation}
\left\{ \overrightarrow{p}\right\} \overset{{\scriptscriptstyle OPT}}{\gets}\underset{\left\{ \overrightarrow{p}\right\} }{\arg\max}\left(\underset{\forall u^{*}\in\left\{ u_{\text{\textsc{tx}}}\right\} }{\sum}\underset{\forall i}{\sum}\widetilde{\mathbb{O}}\left(\widetilde{\mathbb{I}}_{u^{*},i},\,\overrightarrow{p_{i}}_{\left\langle u^{*},u_{\text{\textsc{rx}}}\right\rangle }\right)\right)\label{eq:WPTnoPaths}
\end{equation}
I.e., the maximization of the aggregate output of all paths leading
to the studied receiver, $u_{\text{\textsc{rx}}}$, where $i$ indexes
multiple paths for the same transmitter-receiver pair. Since the objective
is the maximization of power, only the $\mathbf{P}$ attribute of
$\widetilde{\mathbb{O}}$ needs to be retained (cf. def. (\ref{eq:EoutReplace}),
(\ref{eq:EoutReplaceFW})). Moreover, the output can be expressed
as the product of all gain metrics $g_{h}$ (defined in Section~\ref{subsec:Modeling-Core-Functions})
over the tiles comprising each path, multiplied by the input power
of the path:
\begin{equation}
\left\{ \overrightarrow{p}\right\} \overset{{\scriptscriptstyle OPT}}{\gets}\underset{\left\{ \overrightarrow{p}\right\} }{\arg\max}\left(\underset{\forall u^{*}\in\left\{ u_{\text{\textsc{tx}}}\right\} }{\sum}\underset{\forall i}{\sum}\text{\ensuremath{\mathbf{P}}}_{u^{*},i}\cdot\underset{\forall h\in\overrightarrow{p_{i}}_{\left\langle u^{*},u_{\text{\textsc{rx}}}\right\rangle }}{\prod g_{h}}\right)\label{eq:WPT}
\end{equation}
It is noted that the path-centric formulation allows for the employing
the $\prod g_{h}$ expression in relation~(\ref{eq:WPT}). In other
words, treating a path as input to the optimization, allows for calculating
(and caching) its total effect on the transmitter power. An alternative
$f_{h}$-centric formulation would not allow this simple expression,
since the output of a tile\textendash and, thus, its gain\textendash depends
on the input wave and the deployed tile function (e.g., cf. rel.(\ref{eq:FstrUnint})).
Thus, a function-centric formulation would have to follow expression~(\ref{eq:pathO}),
where each gain would receive as input the preceding gains and be
passed as argument to the next one, i.e., $g\left(g\left(g\left(\ldots\right)\right)\right)$.

\subsubsection*{QoS optimization.}

QoS refers to optimizing some aspect of the communication channel
between two users. Without loss of generality, we will focus on the
maximization of the useful-signal-to-interference ratio ($\nicefrac{\ensuremath{\mathbf{P}^{TOT}}}{\ensuremath{\mathbf{I}}}$)
at the studied receiver, $u_{\text{\textsc{rx}}}$. The useful received
signal is expressed as:

\begin{equation}
\ensuremath{\mathbf{P}^{TOT}}=\underset{\forall u^{*}\in\left\{ u_{\text{\textsc{tx}}}\right\} }{\sum}\underset{\forall i}{\sum}\text{\ensuremath{\mathbf{P}}}_{u^{*},i}\cdot\underset{\forall h\in\overrightarrow{p_{i}}_{\left\langle u^{*},u_{\text{\textsc{rx}}}\right\rangle }:\mathcal{P}'}{\prod g_{h}}\label{eq:PtotPDP}
\end{equation}
where
\begin{multline}
\mathcal{P}':\{\overrightarrow{p_{i}}_{\left\langle u^{*},u_{\text{\textsc{rx}}}\right\rangle }\in\mathcal{P}\vert\text{\textsc{Delay}\ensuremath{\left(\overrightarrow{p_{i}}_{\left\langle u^{*},u_{\text{\textsc{rx}}}\right\rangle }\right)}}-\\
\min\left(\left\{ \text{\textsc{Delay}}\left(\overrightarrow{p}\in\mathcal{P}\right)\right\} \right)<\text{D}_{\text{\textsc{th}}}\}\label{eq:PDPrestrict}
\end{multline}
\begin{multline}
\mathcal{P}:\{\overrightarrow{p_{i}}_{\left\langle u^{*},u_{\text{\textsc{rx}}}\right\rangle }\vert\text{\textsc{Tx}\ensuremath{\left(\text{\textsc{LastLink}}\left(\overrightarrow{p_{i}}_{\left\langle u^{*},u_{\text{\textsc{rx}}}\right\rangle }\right)\right)}=}u^{*},\\
\text{\textsc{Rx}\ensuremath{\left(\text{\textsc{FirstLink}}\left(\overrightarrow{p_{i}}_{\left\langle u^{*},u_{\text{\textsc{rx}}}\right\rangle }\right)\right)}=}u_{\text{\textsc{rx}}}\}\label{eq:trrxlabelrestrict}
\end{multline}
Restriction (\ref{eq:trrxlabelrestrict}) (defining path set $\mathcal{P}$)
states that a path carries useful signal : i) if the transmission
is intended for the studied receiver, and ii) is received by a link
labeled to expect the transmitter emission. Therefore, it is implied
that the MIMO capabilities of the user devices have also been configured.
(This is accomplished in Section~\ref{sec:A-K-paths-Approach}).
Restriction~(\ref{eq:PDPrestrict}) defines a subset of $\mathcal{P}$,
based on the latency of each path. Accounting for constructive signal
superposition, the selected paths carrying useful signal must have
bounded latency (i.e., within a time window defined by the fastest
path, plus an application-specific threshold value~$\text{D}_{\text{\textsc{th}}}$).

The interference is expressed similarly to $\ensuremath{\mathbf{P}^{TOT}}$
but over all paths $\overline{\mathcal{P}'}$ leading to the studied
receiver, but not found in~$\mathcal{P}'$:

\begin{equation}
\ensuremath{\mathbf{I}}=\underset{\forall u^{*}\in\left\{ u_{\text{\textsc{tx}}}\right\} }{\sum}\underset{\forall i}{\sum}\text{\ensuremath{\mathbf{P}}}_{u^{*},i}\cdot\underset{\forall h\in\overrightarrow{p_{i}}_{\left\langle u^{*},u_{\text{\textsc{rx}}}\right\rangle }:\overline{\mathcal{P}'}}{\prod g_{h}}\label{eq:Interf}
\end{equation}
Finally, the QoS optimization objective is expressed as:
\begin{equation}
\left\{ \overrightarrow{p}\right\} \overset{{\scriptscriptstyle OPT}}{\gets}\underset{\left\{ \overrightarrow{p}\right\} }{\arg\max}\left(\frac{\ensuremath{\mathbf{P}^{TOT}}}{\ensuremath{\mathbf{I}}}\right)\label{eq:QoSopt}
\end{equation}

\subsubsection*{Eavesdropping mitigation.}

Security concerns pertaining to eavesdropping can be taken into account
and be combined with other objectives~\cite{Liaskos2019ADHOC}. Eavesdropping
mitigation is achieved by ensuring that the employed communication
paths avoid all but the intended user. To this end, consider a 3D
surface, $S\left(u\right)$, around a user $u$. $S\left(u\right)$
can exemplary be a sphere centered at the user's position. We proceed
to discard the paths produced by Algorithm~\ref{alg:PropagationPWE}
that geometrically intersect with the $S\left(u\right)$ (i.e., a
check for intersection between any link in the path and the surface).
Avoiding any user but the intended receiver can then be expressed
as:
\begin{equation}
\overrightarrow{p_{i}}_{\left\langle u^{*},u_{\text{\textsc{rx}}}\right\rangle }\vert\overrightarrow{p_{i}}\cap S\left(u\right)\to\emptyset,\,\forall u\ne u_{\text{\textsc{rx}}},u\in\mathcal{U}\label{eq:SECURITY}
\end{equation}
Restriction~(\ref{eq:SECURITY}) can be further customized, e.g.,
to avoid a targeted set of users only, to discard paths prone to eavesdropping
based on the intersection with some convex hull surface covering many
users, or even to quantify the eavesdropping risk as a scalar value,
such as the minimum distance between a path and a user device.

The expression of a security concern as a restriction over paths,
allows for natural combination between security and QoS or power transfer
objectives. Restriction~(\ref{eq:SECURITY}) can simply filter out
paths before $\mathcal{P}$ (restriction~(\ref{eq:trrxlabelrestrict}))
or after $\mathcal{P}'$ (restriction~(\ref{eq:PDPrestrict})). In
the former case security is prioritized over connectivity, while in
the latter case connectivity comes first.
\begin{rem}
The described eavesdropping mitigation approach naturally limits the
interference caused to unintended recipients. Thus, the eavesdropping
mitigation is not antagonistic to other QoS objectives.
\end{rem}

\subsubsection*{Doppler effect mitigation}

The geometric approach in path restrictions can be extended to mitigate
Doppler effects. Frequency shifts owed to Doppler effects are especially
important in mm-wave communications (and higher frequencies), where
even pedestrian movement rapidly deteriorates the reception quality~\cite{Dopplermillimeter}.
To this end, PWE can strive to keep the last link of communication
paths perpendicular to the trajectory of the user, $\overrightarrow{m}_{u_{\text{\textsc{rx}}}}$,
as follows:
\begin{equation}
\overrightarrow{p_{i}}_{\left\langle u^{*},u_{\text{\textsc{rx}}}\right\rangle }\vert\text{\textsc{LastLink}}\left(\overrightarrow{p_{i}}\right)\bot\overrightarrow{m}_{u_{\text{\textsc{rx}}}}\label{eq:doppler}
\end{equation}
Restriction~(\ref{eq:doppler}) can alternatively be relaxed to a
scalar metric of perpendicularity, e.g., as the inner product of the
unary vector across $\text{\textsc{LastLink}}\left(\overrightarrow{p_{i}}\right)$
and the unary derivative of $\overrightarrow{m}_{u_{\text{\textsc{rx}}}}$
at the user position. Finally, restriction~(\ref{eq:doppler}) can
be combined with all preceding objectives in a manner identical to
the eavesdropping case.

\subsubsection*{User blocking}

Blocking a user from gaining access to another device, such as an
access point, can be facilitated by configuring a PWE to absorb its
NLOS emissions. In this manner, potential security risks can be mitigated
at the physical layer, before expending resources for blocking them
at a higher level (e.g., MAC slots, software authorization steps)~\cite{Liaskos:2018:UAS:3289258.3192336,IEEEcomLiaskos,Liaskos2019ADHOC}.
The mathematical formulation follows naturally from relation~(\ref{eq:WPT}),
by replacing $\max$ with $\min$ within the objective. It is implied
that information on user authorization is passed to the PWE configuration
as an input.

Finally, it is noted that this objective needs \emph{not} be combined
with any other, given that it intends to fully block the physical
connectivity of a user, rather than assign resources to him.

\section{A K-paths Approach for Multi-User Multi-Objective Environment Configuration\label{sec:A-K-paths-Approach}}

The preceding Section formulated the PWE configuration objectives,
using graph paths as inputs. We proceed to define the \emph{\noun{KpConfig}}
heuristic, which configures a PWE for serving a set of user objectives
(Algorithm~\ref{alg:KPATHS}). \emph{\noun{KpConfig }}receives the
following input parameters:
\begin{algorithm}[!t]
\begin{algorithmic}[1]

\Procedure{\noun{KpConfig}}{$\mathcal{G}$,$\left\{ \left\langle u_{\text{\textsc{tx}}},u_{\text{\textsc{rx}}},\text{\textsc{obj}}\right\rangle \right\} $,
$\left\{ \left\langle u_{\text{\textsc{tx}}},\left\{ \widetilde{\mathbb{I}}_{h},h:l_{u,h}\in\left\{ \mathcal{L}_{u}:u=u_{\text{\textsc{tx}}}\right\} \right\} \right\rangle \right\} $}

\State ${\scriptstyle blocked\_pairs}$, ${\scriptstyle sorted\_pairs}$,
$N_{e}$, $K_{e}\gets$\noun{MDFPolicy}( $\mathcal{G}$,$\left\{ \left\langle u_{\text{\textsc{tx}}},u_{\text{\textsc{rx}}},\text{\textsc{obj}}\right\rangle \right\} $);

\State $paths\_rem\gets0;$

\State \textbf{for }$e:\left\langle u_{\text{\textsc{tx}}},u_{\text{\textsc{rx}}},\text{\textsc{obj}}\right\rangle $\textbf{
in} $sorted\_pairs$

\State ~~$paths\gets\left\{ \emptyset\right\} $;

\State ~~\textbf{for} $i=1:1:K_{e}$\textcolor{black}{\State ~~~~$\mathcal{L}_{a}\gets\text{\textsc{FilterLinksByObj}\ensuremath{\left(\mathcal{G},\text{\textsc{obj}}\right)}};$}

\State ~~\textbf{~~}$\overrightarrow{p}{\scriptscriptstyle \gets}\text{\textsc{FindComplexPath}}\left(\mathcal{G},u_{\text{\textsc{tx}}},u_{\text{\textsc{rx}}},\mathcal{L}_{a},paths\right)$;

\State ~~\textbf{~~if $\overrightarrow{p}=\emptyset$}

\State ~~\textbf{~~~~break};\textbf{ }

\State ~~\textbf{~~end if }

\State ~~~~$paths\gets paths+\overrightarrow{p}$;

\State ~~\textbf{end for}

\State ~~\textbf{if} $\left\Vert paths\right\Vert =0$

\State ~~~~\textbf{continue}; \emph{//pair is disconnected.}

\State ~~\textbf{end}

\State~~$max\_paths\gets\min\left\{ N_{e}+paths\_rem,\left\Vert paths\right\Vert \right\} $;

\State ~~$paths\gets\text{\textsc{\ensuremath{{\scriptstyle \text{\textsc{FilterPathsByObj}}}}}}\left(paths,max\_paths,\text{\textsc{obj}}\right)$;

\State ~~$paths\_rem{\scriptstyle \gets}paths\_rem{\scriptstyle +}{\scriptstyle \max}\left\{ {\scriptstyle max\_paths-\left\Vert paths\right\Vert ,0}\right\} $;

\State ~~$\text{\textsc{Deploy}}\left(\mathcal{G},paths\right)$;

\State \textbf{end for}

\State $\text{\textsc{DeployBlocks}}\left(\mathcal{G},blocked\_pairs\right)$;

\State $f_{h}\gets f_{h}^{ABS},\,\forall h\in\mathcal{G}:f_{h}=\emptyset$;

\EndProcedure

\end{algorithmic}

\caption{\label{alg:KPATHS}The K-Paths approach for configuring PWEs.}
\end{algorithm}
\begin{itemize}
\item The PWE graph $\mathcal{G}$ comprising the sub-graphs of connected
users, $\mathcal{G}\left\langle \mathcal{U},\mathcal{L}_{u}\right\rangle $,
and connectable tiles, $\mathcal{G}\left\langle \mathcal{H},\mathcal{L}_{h}\right\rangle $.
We consider the latter as static and, therefore, pre-processable.
Particularly, we assume that a custom number of node-disjoint shortest
paths can be pre-calculated and cached, for each tile pair. The link
weight for such calculations is the link \noun{Delay}, and the ensuing
paths are considered to be filtered based on their total steering
gain (cf. the $\prod g_{h}$ expression in relation~(\ref{eq:WPT}))
and a custom acceptable threshold. Subsequent calls to well-known
path finding algorithms (e.g., \noun{ShortestPath}, \noun{KShortestPaths}~\cite{brander1996comparative})
are considered to be executed on top of this cache.
\item The set of communicating user pairs and their objectives. Multicast
groups are expressed as multiple pairs with the same transmitter.
The objective \noun{obj} is a set of binary flags, each denoting a
type from Section~\ref{subsec:Modeling-connectivity-objectives}.
We assume that symmetric pairs have been filtered out of this parameter.
Complimenting Remark~\ref{rem:F_symmetry-}, a pair $\left\langle u_{\text{\textsc{tx}}},u_{\text{\textsc{rx}}},\text{\textsc{obj}}\right\rangle $
is symmetric to $\left\langle u_{\text{\textsc{rx}}},u_{\text{\textsc{tx}}},\text{\textsc{obj}}\right\rangle $,
highlighting that the objective must also be the same. Moreover, disconnected
pairs (i.e., without any \noun{ShortestPath} in $\mathcal{G}$) are
also considered as filtered out.
\item The inputs $\widetilde{\mathbb{I}}_{h}$ and the affected tiles, for
each transmitter.
\end{itemize}
The use of the link \noun{Delay} as the link weight prioritizes shorter
links to assemble a path. In practice, the output of a tile function
may digress from the intended as the distance from the tile increases~\cite{CUIcoding2018,MSSurveyAllFunctionsAndTypes}.
In this aspect, shorter links may favor a more consistent behavior.
Additionally, the output of \emph{\noun{KpConfig}} is a set of paths
per pair (which correspond to tile functions as described in Section~\ref{subsec:Modeling-connectivity-objectives}).
Thus, the \noun{Tx} and \noun{Rx} labels per user link are naturally
produced by \emph{\noun{KpConfig,}} without further steps. \\
\begin{algorithm}[t]
\begin{algorithmic}[1]

\Procedure{${\scriptstyle blocked\_pairs}$, ${\scriptstyle sorted\_pairs}$,
$N_{e}$, $K_{e}$\noun{: }\\
\noun{MDFPolicy}}{ $\mathcal{G}$,$\left\{ \left\langle u_{\text{\textsc{tx}}},u_{\text{\textsc{rx}}},\text{\textsc{obj}}\right\rangle \right\} $}

\State $n_{u}\gets\left\{ 0,\ldots,0\right\} $; \emph{//for all
users.}

\State $sorted\_pairs\gets\left\{ \emptyset\right\} $;

\State $blocked\_pairs\gets\left\{ \emptyset\right\} $;

\State $v_{e}\gets\left\{ \emptyset\right\} $;

\State $K_{e}\gets\left\{ \emptyset\right\} $;

\State $N_{e}\gets\left\{ \emptyset\right\} $;

\State \textbf{for }$e:\left\langle u_{\text{\textsc{tx}}},u_{\text{\textsc{rx}}},\text{\textsc{obj}}\right\rangle $\textbf{
in} $\left\{ \left\langle u_{\text{\textsc{tx}}},u_{\text{\textsc{rx}}},\text{\textsc{obj}}\right\rangle \right\} $

\State ~~\textbf{if} $\text{\textsc{obj}}=\text{\textsc{block}}$//
\emph{$u_{\text{\textsc{rx}}}$ is arbitrary.}

\State ~~\textbf{~~$blocked\_pairs\gets blocked\_pairs+e$;}

\State ~~\textbf{~~continue;}

\State ~~\textbf{else;}

\State ~~\textbf{~~$sorted\_pairs\gets sorted\_pairs+e$;}

\State ~~\textbf{end if}

\State ~~$K_{e}\gets\min\left\{ \left\Vert \mathcal{L}_{u:u_{\text{\textsc{tx}}}}\right\Vert ,\left\Vert \mathcal{L}_{u:u_{\text{\textsc{rx}}}}\right\Vert \right\} ;$

\State ~~$\left\{ \overrightarrow{p}\right\} _{e}\gets\text{\textsc{KShortestPaths}}\left(\mathcal{G},K_{e},u_{\text{\textsc{tx}}},u_{\text{\textsc{rx}}}\right)$;

\State ~~$v_{e}\gets\text{\textsc{Mean}}\left(\left\{ \text{\textsc{Delay}}\left(\overrightarrow{p}\right)\right\} _{e}\right)$;

\State ~~$n_{u_{\text{\textsc{tx}}}}\gets n_{u_{\text{\textsc{tx}}}}+\left\Vert \left\{ \overrightarrow{p}\right\} _{e}\right\Vert $;

\State ~~$n_{u_{\text{\textsc{rx}}}}\gets n_{u_{\text{\textsc{rx}}}}+\left\Vert \left\{ \overrightarrow{p}\right\} _{e}\right\Vert $;

\State \textbf{end for}

\State $sorted\_pairs\gets\text{\textsc{SortDesc}}\left(sorted\_pairs,v_{e}\right)$;

\State \textbf{for }$e:\left\langle u_{\text{\textsc{tx}}},u_{\text{\textsc{rx}}},\text{\textsc{obj}}\right\rangle $\textbf{
in} $sorted\_pairs$

\State  ~~$N_{e}\gets\max\left\{ \min\left\{ \left\lfloor \frac{\left\Vert \mathcal{L}_{u:u_{\text{\textsc{tx}}}}\right\Vert }{n_{u_{\text{\textsc{tx}}}}}\right\rfloor ,\left\lfloor \frac{\left\Vert \mathcal{L}_{u:u_{\text{\textsc{rx}}}}\right\Vert }{n_{u_{\text{\textsc{rx}}}}}\right\rfloor \right\} ,1\right\} $;

\State  \textbf{end for }

\EndProcedure

\end{algorithmic}

\caption{\label{alg:MDFra}The ``Most Distant Pair First'' resource sharing
policy for PWEs.}
\end{algorithm}
Since \emph{\noun{KpConfig }}seeks to serve multiple user pairs, a
resource sharing policy needs to be employed in the general case.
Thus, at line 2, \emph{\noun{KpConfig,}}\emph{ }calls upon the 'Most-Distant-First'
(\noun{MDFPolicy}) subroutine, which is an exemplary policy.

\noun{MDFPolicy}, formulated as Algorithm~\ref{alg:MDFra}, primarily
returns a sorting of the user pairs ($sorted\_pairs$) by descending
priority order, as well as the number of paths to allocate per pair
$e$, $N_{e}$. At lines $9-21$, \noun{MDFPolicy} filters out any
pairs with user access \noun{block} objectives, since these will not
require connecting paths. Remaining pairs are sorted by descending
average delay, calculated over $K_{e}$ shortest paths in graph $\mathcal{G}$,
$K_{e}$ being the minimum number of user links in the pair $e$.
This simple heuristic intends to prioritize distant pairs, on the
grounds of experiencing a lesser degree of propagation control. Finally,
the tentative number of paths, $N_{e}$, to allocate pair pair is
returned at line $24$, as the minimum ratio of number of user links
divided by the number of pairs this user belongs to.

\emph{\noun{KpConfig (}}Algorithm~\ref{alg:KPATHS}\emph{\noun{)
}}then resumes its operation. In general, it comprises an exploratory
evaluation of $K_{e}$ paths per sorted pair (lines $6-13$), eventually
keeping and deploying at most $N_{e}$ subject to compliance with
objectives (lines $17-20$). Left-over path allocations are redistributed
to ensuing pairs via the $paths\_rem$ variable (lines $3,19$).\\
\begin{algorithm}[t]
\begin{algorithmic}[1]

\Procedure{$\mathcal{L}_{a}:\text{\textsc{FilterLinksByObj}}$}{$\mathcal{G},\text{\textsc{obj}}$}

\State $\mathcal{L}_{a}\gets\left\{ \right\} ;$

\State \textbf{switch} $\text{\textsc{obj}}$:

\State ~~\textbf{case} {\small{}$\text{\textsc{MitigateEavesDrop}}$:
}\emph{\small{}//Inputs of (\ref{eq:SECURITY}) implied.}{\small\par}

\State ~~~~$\mathcal{L}_{a}\gets\mathcal{L}_{a}+\text{relation\,(\ref{eq:SECURITY})};$

\State ~~\textbf{case} {\small{}$\text{\textsc{MitigateDoppler}}$:
}\emph{\small{}//Inputs of (\ref{eq:doppler}) implied.}{\small\par}

\State ~~~~$\mathcal{L}_{a}\gets\mathcal{L}_{a}+\text{relation\,(\ref{eq:doppler})};$

\State \textbf{end switch}

\EndProcedure

\end{algorithmic}

\caption{\label{alg:FiltLinksByObj}Obtaining links in conflict with eavesdropping
and Doppler effect mitigation objectives.}
\end{algorithm}
The exploration of $K_{e}$ paths at lines $6-13$ is also a point
for enforcing eavesdropping and Doppler effect mitigation objective
types. At line $7$, the \noun{FilterLinksByObj} subroutine is called
(Algorithm~\ref{alg:FiltLinksByObj}), which returns the links $\mathcal{L}_{a}$
of $\mathcal{G}$ that are in conflict with the objectives. The use
of $\texttt{switch}$ without $\texttt{break}$ statements implies
that both objectives can be active for the same user pair. \\
\begin{algorithm}[t]
\begin{algorithmic}[1]

\Procedure{$\overrightarrow{p}$\noun{: FindComplexPath}}{ $\mathcal{G},u_{\text{\textsc{tx}}},u_{\text{\textsc{rx}}},\mathcal{L}_{a},paths$}

\State $\mathcal{G}'\gets\mathcal{G}\left\langle \left\{ \mathcal{H},u_{\text{\textsc{tx}}},u_{\text{\textsc{rx}}}\right\} ,\left\{ {\scriptstyle \mathcal{L}_{h}},{\scriptstyle \mathcal{L}_{u_{\text{\textsc{tx}}}}},{\scriptstyle \mathcal{L}_{u_{\text{\textsc{rx}}}}}\right\} -{\scriptstyle \mathcal{L}_{a}}-{\scriptstyle paths}\right\rangle $;

\State $\mathcal{G}^{*}\gets\mathcal{G}'-\left\langle \left\{ \mathcal{H}\vert f_{h}=\emptyset\right\} \right\rangle $;

\State $\overrightarrow{p}\gets\text{\textsc{ShortestPath}}\left(\mathcal{G}^{*},u_{\text{\textsc{tx}}},u_{\text{\textsc{rx}}}\right)$;

\State \textbf{if} $\overrightarrow{p}\ne\emptyset$

\State ~~\textbf{return} $\overrightarrow{p}$; \emph{//Path without
deployed functions found. }

\State \textbf{end} \textbf{if}

\State $\overrightarrow{p}^{*}\gets\text{\textsc{ShortestPath}}\left(\mathcal{G}',u_{\text{\textsc{tx}}},u_{\text{\textsc{rx}}}\right)$;

\State \textbf{if} $\overrightarrow{p}^{*}=\emptyset$

\State ~~\textbf{return} $\emptyset$; \emph{//No path exists}

\State \textbf{end} \textbf{if}

\State $\overrightarrow{p}\gets\left\{ \right\} $; $h'\gets\emptyset$;

\State \textbf{for} $\left\langle l,h\right\rangle $ \textbf{in}
$\overrightarrow{p}^{*}$

\State ~~\textbf{if} $f_{h}=\emptyset$

\State ~~~~$\overrightarrow{p}\gets\overrightarrow{p}+l$;

\State ~~\textbf{else}

\State ~~~~$\left\langle l',h'\right\rangle \gets f_{h}\left(l\right)$;

\State ~~~~$\overrightarrow{p}\gets\overrightarrow{p}+l'$;

\State ~~~~\textbf{break};

\State ~~\textbf{end} \textbf{if}

\State \textbf{end} \textbf{for}

\State $\overrightarrow{p}^{*}\gets\text{\textsc{Self}\ensuremath{\left(\mathcal{G},h',u_{\text{\textsc{rx}}},\mathcal{L}_{a},paths-\overrightarrow{p}\right)}}$;

\State \textbf{if} $\overrightarrow{p}^{*}=\emptyset$

\State ~~\textbf{return} $\emptyset$;

\State \textbf{else}

\State ~~\textbf{return} $\overrightarrow{p}+\overrightarrow{p}^{*}$;

\State \textbf{end} \textbf{if}

\EndProcedure

\end{algorithmic}

\caption{\label{alg:ComplexPath}The process for finding a single path in a
PWE.}
\end{algorithm}
At line $7$ of \emph{\noun{KpConfig (}}Algorithm~\ref{alg:KPATHS}\emph{\noun{)}},
the links $\mathcal{L}_{a}$ are passed to the \noun{FindComplexPath}
subroutine (Algorithm~\ref{alg:ComplexPath}), which seeks to find
a path in $\mathcal{G}$ that connects the pair while avoiding the
links $\mathcal{L}_{a}$ and the links over the already deployed $paths$.
Two subgraphs are created at lines $2$ and $3$ of Algorithm~\ref{alg:ComplexPath}.
The subgraph $\mathcal{G}'$, which removes the aforementioned links
from the original graph $\mathcal{G}$, and $\mathcal{G}^{*}$ which
also removes the already configured tiles (nodes) of $\mathcal{G}'$.
\noun{FindComplexPath }first tries to find a connecting path in $\mathcal{G}^{*}$
(line~$4$). If found, the path comprises unconfigured tiles only,
thus being in compliance with Remark~\ref{rem:DONOTREUSE} and avoiding
the combination of tile functions. If not found, the search continues
on $\mathcal{G}'$. A found path is then bound to contain already
configured tiles. Lines $13-21$ iterate over the path links and tiles
until the first configured tile is found (lines $16-20$). Then, the
output (link $l'$ and reached tile $h'$) of the already deployed
function are calculated (line $17$). The subroutine is then recursively
called to find a path from $h'$ to the intended receiver, while also
avoiding links already visited.\\
\begin{algorithm}[t]
\begin{algorithmic}[1]

\Procedure{\noun{$sel\_paths$:}\\
\noun{ FilterPathsByObj}}{$paths$,$max\_paths$\emph{,}\emph{\noun{obj}}}

\State \textbf{switch} $\text{\textsc{obj}}$

\State ~~\textbf{case }$\text{\textsc{MaxPower}}$:\textbf{ }\textcolor{black}{\emph{\footnotesize{}//Inputs
of (\ref{eq:WPT}) implied.}}{\footnotesize\par}

\State ~~~~$paths\gets\text{\textsc{SortDesc}}\left(paths\right)$;
\textcolor{black}{\emph{\footnotesize{}// by $\text{\ensuremath{\mathbf{P}}}\cdot\prod g_{h}$.}}{\footnotesize\par}

\State ~~~~$paths\gets\text{\textsc{KeepTopN}}\left(paths,max\_paths\right)$;

\State ~~\textbf{~~break};\textbf{ }

\State ~~\textbf{case }$\text{\textsc{MaxSIR}}$:\textbf{ }\textcolor{black}{\emph{\footnotesize{}//Inputs
of (\ref{eq:PtotPDP}) implied.}}\textbf{}

\State ~~~~$paths\gets\text{\textsc{SortAsc}}\left(paths\right)$;
\textcolor{black}{\emph{\footnotesize{}// by path $\text{\textsc{Delay}}$.}}{\footnotesize\par}

\State ~~~~~~\textbf{for} i=1:1:$\left\Vert paths\right\Vert $

\State ~~~~~~~~$\mathbf{P}_{i}^{TOT}\gets$~~~~~\textcolor{black}{\emph{\footnotesize{}//by
rel. (\ref{eq:PtotPDP}),(\ref{eq:PDPrestrict}).}}{\footnotesize\par}

\State ~~~~~~~~\textbf{~~$\mathbf{P}^{TOT}\left(paths_{i,\min\left\{ i+max\_paths,\left\Vert paths\right\Vert \right\} }\right)$ }

\State ~~~~~~\textbf{end for}

\State ~~~~~~\textbf{$i^{OPT}\gets arg\max\left\{ \mathbf{P}_{i}^{TOT}\right\} $};

\State ~~~~~~$paths_{i^{OPT},\min\left\{ i^{OPT}+max\_paths,\left\Vert paths\right\Vert \right\} }$;

\State ~~\textbf{~~break};\textbf{ }

\State \textbf{end switch}

\State $sel\_paths\gets paths$;

\EndProcedure

\end{algorithmic}

\caption{\label{alg:SELBYOBJ}Objective-oriented propagation path selection.}
\end{algorithm}
Returning to the workflow of\emph{\noun{ KpConfig (}}Algorithm~\ref{alg:KPATHS}),
line $17$ is reached with a maximum of $K_{e}$ found paths for the
studied pair. After accounting for any surplus paths that have remained
from preceding pairs (line~$17$), an objective-driven selection
of paths takes place at line~$18$ via the \noun{FilterPathsByObj}
subroutine (Algorithm~\ref{alg:SELBYOBJ}).

\noun{FilterPathsByObj} considers a power maximization \emph{or }a
QoS optimization objective, since these cannot be generally combined.
The path selection for power transfer maximization is straightforward:
the paths are sorted by total power and the top ones are selected.
For a QoS optimization, the paths are first order by total delay.
Then, a sliding window-based selection takes place, keeping the path
subset that maximizes relation \textcolor{black}{(\ref{eq:PtotPDP})
subject to (\ref{eq:PDPrestrict}). Notice this approach does not
take into account the signal interference (relation~(\ref{eq:Interf})).
Instead, an eavesdropping objective can be added to all user pairs,
thus naturally minimizing or mitigating interference as well. }

\emph{\noun{KpConfig (}}Algorithm~\ref{alg:KPATHS}) proceeds to
\noun{Deploy} the selected paths at line~$20$, updating the $f_{h}$
status in $\mathcal{G}$ as well. The deployment takes place by setting
collimation functions, $f_{h}^{\text{\textsc{Col}}}$, at the first
and last tile of the path, and steering functions, $f_{h}^{\text{\textsc{Str}}}$,
to intermediate ones. Already configured tiles are not affected. Required
modifiers, $m_{h}^{\text{\textsc{Pha}}}$ or $m_{h}^{\text{\textsc{Pol}}}$
are applied afterwards sequentially, at the tile that yields the smallest
effect on the total path gain $\prod g$ .
\begin{algorithm}[t]
\begin{algorithmic}[1]

\Procedure{$\textsc{DeployBlocks}$}{$\mathcal{G},blocked\_pairs$}

\State \textbf{for }$e:\left\langle u_{\text{\textsc{tx}}},u_{\text{\textsc{rx}}},\text{\textsc{obj}}\right\rangle $\textbf{
in} $blocked\_pairs$

\State ~~\textbf{~~for }$l_{u,h^{*}}\in\mathcal{L}_{u}\vert u=u_{\text{\textsc{tx}}}$

\State ~~\textbf{~~}$\overrightarrow{p}\gets\text{\textsc{FindComplexPath}}\left(\mathcal{G},h^{*},u_{\text{\textsc{rx}}}\right)$;

\State ~~\textbf{~~~~for }$h\in\overrightarrow{p}$

\State ~~\textbf{~~~~~~if }$f_{h}=\emptyset$\textbf{ }

\State ~~\textbf{~~~~~~~~}$f_{h}\gets f_{h}^{ABS}$;

\State ~~\textbf{~~~~~~~~break;}

\State ~~~~~~\textbf{~~end if}

\State ~~\textbf{~~~~end for}

\State ~~\textbf{~~end for}

\State \textbf{end for}

\EndProcedure

\end{algorithmic}

\caption{\label{alg:DeployBlocks}The process for blocking user access in PWE.}
\end{algorithm}
At line $22$, \emph{\noun{KpConfig (}}Algorithm~\ref{alg:KPATHS})
takes action to block unauthorized users, contained in the $blocked\_pairs$
set. The process is given in the subroutine \noun{DeployBlocks} (Algorithm~\ref{alg:DeployBlocks}).
Since the objective is to block a single user from all others, we
proceed to iterate over the user links of $u_{\text{\textsc{tx}}}$
and fully absorb the emissions at each one. Notice that the immediately
affected tiles may have a deployed function. Thus, in the general
case, \noun{DeployBlocks} searches for a path from $u_{\text{\textsc{tx}}}$
to an arbitrary node (random user $u_{\text{\textsc{rx}}}$ or any
random tile). Once found, the path is iterated over and a full absorption
function, $f_{h}^{\text{\textsc{Abs}}}$, is deployed at the first
unconfigured tile $h$.

Finally, \emph{\noun{KpConfig (}}Algorithm~\ref{alg:KPATHS}) concludes
at line~$23$ by setting all unused tiles in $\mathcal{G}$ to absorb
from an arbitrary direction (e.g., from the tile surface normal).
Thus, parasitic function outputs within PWE can be attenuated. It
is noted that this step can be omitted, e.g., in cases where limiting
the total number of configured tiles is a concern.

\section{Evaluation\label{sec:Evaluation}}

We proceed to evaluate the performance of \noun{KpConfig} in a set
of floorplans, users and objectives. We seek to demonstrate and visualize
compliance to performance objectives, and deduce the performance gains
in comparison with natural propagation (non-PWE).

The evaluation is based on a novel tool, implementing the process
described in Section~\ref{sec:A-Graph-based-Model}, developed specifically
for simulating PWEs. The tool receives a floorplan, a set of configured
tiles, as well as users locations and user transmission characteristics.
Subsequently, it essentially executes the process of Algorithm~\ref{alg:PropagationPWE}.
Its simulation output is the power-delay profile for each user, which
is then used for quantifying the compliance with the set objectives.
The tool is implemented in JAVA over the AnyLogic platform~\cite{XJTechnologies.2013}.
The latter is chosen due to its strong visualization capabilities,
and its versatility in combining simulation models (discrete events,
agents, continuous processes) to describe a complex system. The developed
tool, freely available on demand, allows for realistic simulation
of PWE, supporting the following features:
\begin{itemize}
\item Using any EM tile profile as input, supporting any physical implementation
approach. The EM profiles can be input from Computational EM packages,
or real measurements in a well-defined format.
\item Customized antenna radiation patterns with MIMO support.
\item Allowing for controllable reflection, refraction and diffraction of
EM waves, along with any described metasurface functionality.
\item Flexible scenario creation via a 3D Graphical User Interface, allowing
for varying tile topologies, dimensions, floorplan, user placement
and roles, user mobility trajectories, and partially coated environments.
\item Flexible user objective definition mechanism, readily supporting all
described objectives and allowing for creating custom ones.
\item Multi-cast and broadcast support.
\item Script-able and automate-able, parallelized simulations, allowing
for automated parameter optimization, parameter sensitivity, Monte
Carlo experiments, and general-purpose heuristic optimization runs~\cite{XJTechnologies.2013}.
\begin{table}[t]
\centering{}\caption{\textsc{\label{tab:TSimParams}Persistent Simulation parameters.}}
\begin{tabular}{|c|c|}
\hline
Ceiling Height & $3$~$m$\tabularnewline
\hline
Tile Dimensions  & \textbf{$1\times1\,m$} \tabularnewline
\hline
Tile Functions & $f_{h}^{\text{\textsc{Col}}}$, $f_{h}^{\text{\textsc{Str}}}$, $f_{h}^{\text{\textsc{Abs}}}$\tabularnewline
\hline
Non-HSF surfaces & Perfect reflectors\tabularnewline
\hline
User scattering model & Blocking spheres, radius $0.5\,m$\tabularnewline
\hline
Frequency & $2.4\,GHz$ \tabularnewline
\hline
Tx Power & $-30\,dBm$\tabularnewline
\hline
\multirow{2}{*}{Antenna type} & Single $a^{o}$-lobe sinusoid,\tabularnewline
 &  pointing at $\phi^{o},\theta^{o}$ (cf. Fig.~\ref{fig:USERSETUP})\tabularnewline
\hline
Max ray bounces & $50$\tabularnewline
\hline
Min considered ray power & -$250$~dBm\tabularnewline
\hline
\end{tabular}
\end{table}
\begin{figure}[t]
\begin{centering}
\includegraphics[width=0.4\columnwidth]{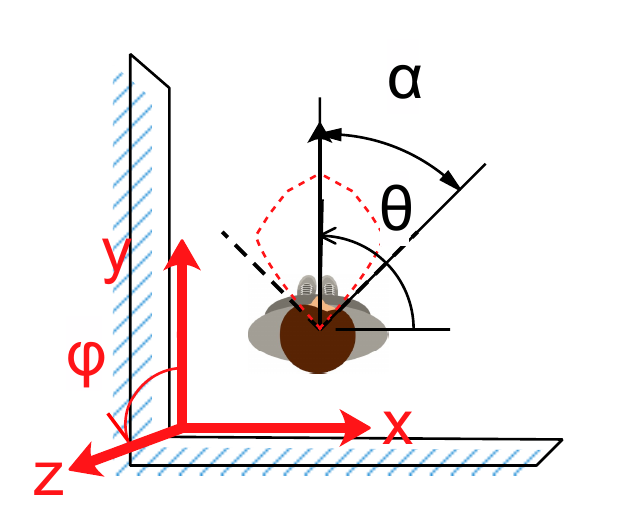}
\par\end{centering}
\caption{\label{fig:USERSETUP}User location and beam orientation coordinate
system. The origin is at the lower left corner of each floorplan. }
\end{figure}
\end{itemize}
Table~\ref{tab:TSimParams} summarizes the persistent parameters
across all subsequent tests. The following notes are made:
\begin{itemize}
\item The user positions and radiation characteristics are considered to
be known. These can be deduced by a user localization scheme specific
for PWEs~\cite{NANOCOMlocalizeSDM}, or a third party scheme.
\item All tiles have the same EM profile (cf.~Remark~\ref{rem:PROFILE}),
which is defined as follows: all impinging waves are attenuated by
a constant factor of $g=1$~\% of the carried power, for any type
of deployed function. While simplified, this profile can represent
existing EM models~\cite{CUIcoding2018}, when the meta-atom resolution
is infinite~\cite{wu2018intelligent}.
\item The considered tile functionalities include collimation, steering
and absorption, while modifiers are not taken into account. As previously
described, collimation is the effect of aligning EM waves to propagate
over a flat front, rather than to dissipate over an ever-growing sphere.
Thus, the path loss between two tiles in a PWE is not subject to the
$\propto\nicefrac{1}{d^{2}}$ rule, $d$ being their distance (cf.~Fig.~\ref{fig3})~\cite{Liaskos2019ADHOC}.
This rule is only valid for the first impact, i.e., from the transmitter
to its LOS tiles. The antenna aperture effect and gain are taken into
account as usual.
\item The antenna patterns of the transmitter and the receiver are simplified
as single-lobe sinusoids, with the characteristics and $\theta,\,\phi$
orientation shown in Fig.~\ref{fig:USERSETUP} and Table~\ref{tab:TSimParams}.
In some scenarios, we assume that the mobile devices have beamforming
capabilities and are able to turn the antenna lobe towards the ceiling,
in conjunction with the mobile devices' gyroscopes.
\end{itemize}
Finally, all walls, ceilings and floors are considered as fully coated
with HyperSurfaces, unless otherwisely stated (defined per case).

\subsection{Multi-User Multi-Objective Showcase}

\begin{table}[tbh]
\centering{}\caption{\label{tab:SetupMUMO}Scenario setup for the multi-user multi-objective
showcase.}
\begin{tabular}[b]{cc}
\begin{tabular}{@{}l}
\toprule
\multicolumn{1}{l}{\textsc{User: Position, $\alpha$,$\phi$}}\tabularnewline
\midrule
0: {[}1.0,10.0,1.0{]},10.0$^{o}$,$0^{o}$\tabularnewline
1: {[}8.5,11.0,1.0{]},80.0$^{o}$,$0^{o}$ \tabularnewline
2: {[}15.6,11.5,1.0{]},10.0$^{o}$,$0^{o}$ \tabularnewline
3: {[}11.0,5.4,1.0{]},10.0$^{o}$,$0^{o}$ \tabularnewline
4: {[}6.0,6.0,1.0{]},10.0$^{o}$,$0^{o}$ \tabularnewline
5: {[}8.5,1.6,1.0{]},60.0$^{o}$,$0^{o}$ \tabularnewline
\midrule
\multicolumn{1}{l}{\textsc{Pair: Objective}}\tabularnewline
\midrule
\textcolor{magenta}{0$\to$2} : ${\scriptstyle {\textsc{MaxPower},\textsc{EavesMit}[[\text{\textsc{All}}]]}}$\tabularnewline
\textcolor{blue}{1$\to$3} : ${\scriptstyle {\textsc{MaxPower},\textsc{EavesMit}[[\text{\textsc{All}}]]}}$\tabularnewline
\textcolor{green}{1$\to$4} : ${\scriptstyle {\textsc{MaxSIR},\textsc{EavesMit}[[\text{\textsc{All}}]]}}$\tabularnewline
\textcolor{red}{5$\to$$\times$} : ${\scriptstyle {\textsc{Block}}}$\tabularnewline
\bottomrule
\end{tabular} & %
\begin{tabular}{c}
\tabularnewline
\includegraphics[width=0.4\columnwidth]{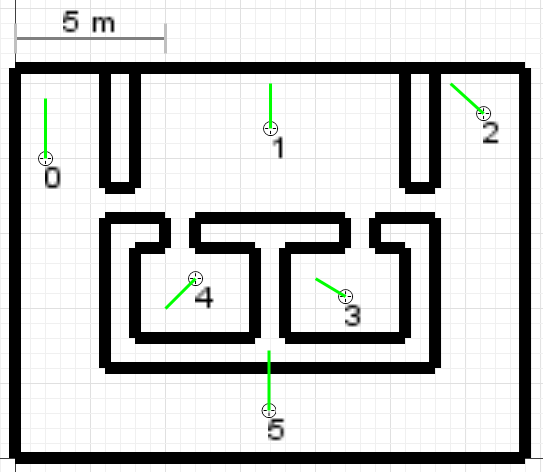}\tabularnewline
\end{tabular}\tabularnewline
\end{tabular}
\end{table}
\begin{figure}[t]
\begin{centering}
\subfloat[\label{fig:PWE}PWE]{\begin{centering}
\includegraphics[width=0.95\columnwidth]{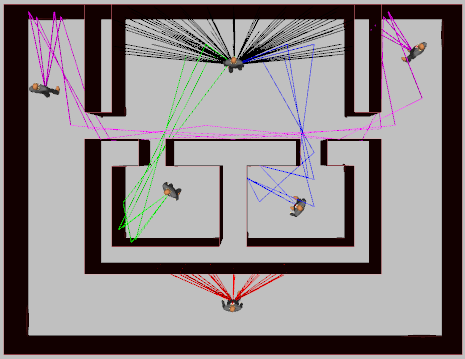}
\par\end{centering}
}
\par\end{centering}
\begin{centering}
\subfloat[\label{fig:No-PWE}No PWE]{\begin{centering}
\includegraphics[width=0.95\columnwidth]{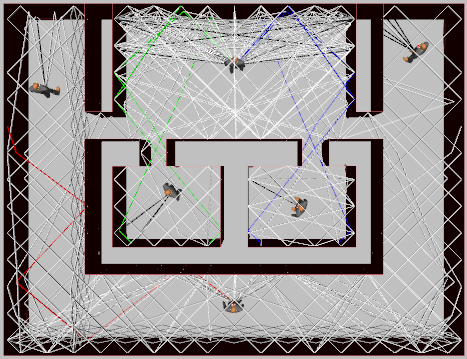}
\par\end{centering}
}
\par\end{centering}
\caption{\label{fig:MUMO_propag}Propagation with and without PWE.}
\end{figure}
This scenario showcases the PWE performance for three communicating
pairs and one unauthorized user in a floorplan, as described Table~\ref{tab:SetupMUMO}.
The pairs have various different objectives, while there is also a
multicast group ($1$ to $3,4$). Moreover, the antenna characteristics
vary across the users.

The achieved wireless propagation, with and without PWE, is illustrated
in Fig.~\ref{fig:MUMO_propag}. The PWE, configured via the proposed
$\text{\textsc{KpConfig}}$, customizes the EM propagation to uphold
all objectives (Fig.~\ref{fig:PWE}). Specifically, the propagation
for pair $0\to2$ follows a short route within the floorplan, to achieve
maximal received power, while also avoiding all potential eavesdroppers.
The pair $1\to3$ is treated similarly, i.e., maximizing the received
powered without QoS concerns stemming from the length of each ray.
For the pair $1\to4$, however, QoS is taken into account, resulting
into a selection of rays that have similar total length, as shown
in Fig.~\ref{fig:PWE}. Notice that, once the multicast group is
served, the environment is tuned to absorb all unemployed emissions
from user~$1$ (black lines). Finally, user~5 has its EM emissions
absorbed, and is thus blocked from accessing all other users.

Natural propagation (i.e., non-PWE) within the same floorplan is shown
in Fig.~\ref{fig:No-PWE}. While some pairs achieve a degree of connectivity
(e.g., $1\to4$), the propagation is expectedly chaotic. The white
lines denote stray rays within the floorplan, which eventually attenuate
and disappear without reaching a user. Notice that the floorplan design,
coupled with the highly directional antenna lobes of some users (e.g.,
$\alpha=10^{o}$) naturally secludes users and hinders connectivity.

\begin{figure}[t]
\begin{centering}
\includegraphics[viewport=0bp 0bp 450bp 130bp,clip,width=1\columnwidth]{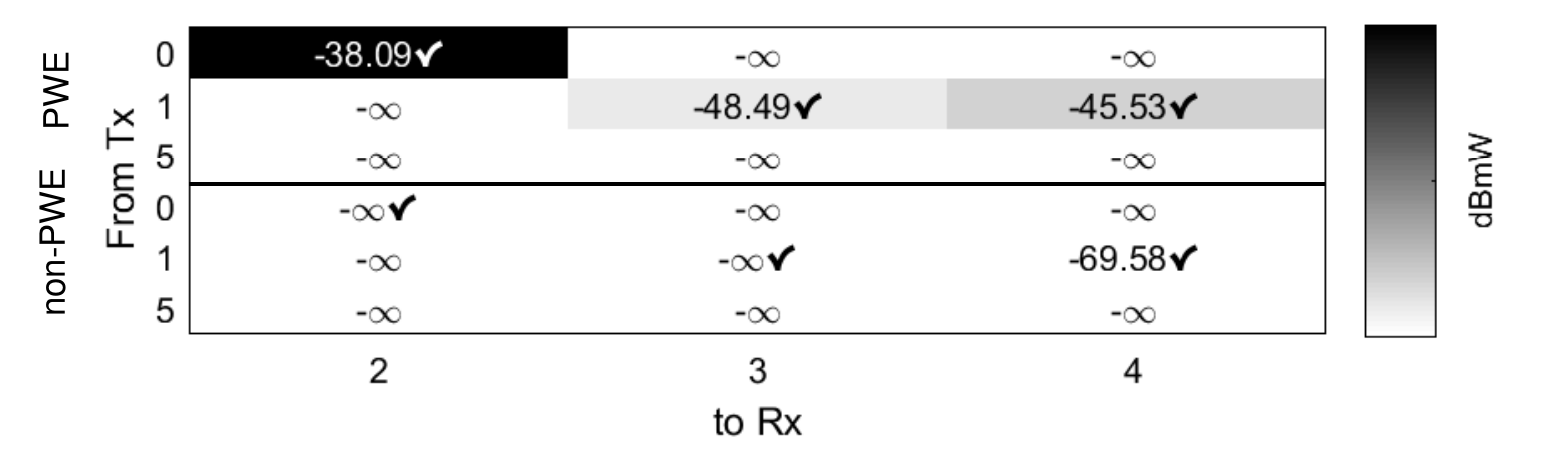}
\par\end{centering}
\caption{\label{fig:MUMOplot}Comparative power reception with and without
PWE. \CheckmarkBold{} denotes intended pair connectivity. }
\end{figure}
The received power levels are shown in Fig.~\ref{fig:MUMOplot}.
PWE achieves objective-compliant connectivity with a relatively small
loss, ranging from $\sim8$ to $15$~dBmW. On the other hand, the
non-PWE case achieves connectivity only for pair $1\to4$, with a
considerably high loss of $\sim30$~dBmW. It is noted that the PWE
results of Fig.~\ref{fig:MUMOplot} are calculated as the total-power-minus-interference.
The classification of a ray as interfering factors for the MIMO labeling
of the user links produced by $\text{\textsc{KpConfig}}$. (The interference
was zero in all cases). In the non-PWE case, the interference classification
does not consider any MIMO labeling, since there is not automatic
way of setting it.

\subsection{Doppler Effect Mitigation Showcase}

\begin{table}[t]
\centering{}\caption{\label{tab:SetupDoppler}Scenario setup for the Doppler effect mitigation
showcase.}
\begin{tabular}[b]{cc}
\begin{tabular*}{3.8cm}{@{\extracolsep{\fill}}@{}l}
\multicolumn{1}{l}{\textsc{User: Position, $\alpha$,$\phi$}}\tabularnewline
\midrule
0: {[}17.0,3.2,1.0{]},30.0$^{o}$,90.0$^{o}$ \tabularnewline
1: {[}2.0,1.7,1.0{]},30.0$^{o}$,90.0$^{o}$ \tabularnewline
\midrule
\multicolumn{1}{l}{\textsc{Pair: Objective}}\tabularnewline
\midrule
\textcolor{cyan}{1$\to$0} : $\{{\scriptstyle \textsc{MaxPower},\textsc{EavMit}[[0]]}$,\tabularnewline
~~$\textsc{DopplerMit}[trajectory]$\}\tabularnewline
\bottomrule
\end{tabular*} & %
\begin{tabular}{c}
\tabularnewline
\includegraphics[width=0.35\columnwidth]{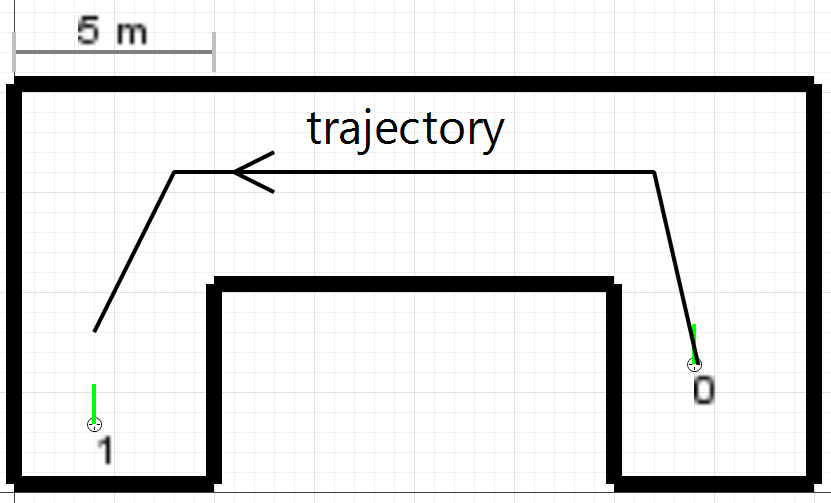}\tabularnewline
\end{tabular}\tabularnewline
\end{tabular}
\end{table}
\begin{figure}[t]
\begin{centering}
\includegraphics[viewport=0bp 0bp 450bp 400bp,clip,width=1\columnwidth]{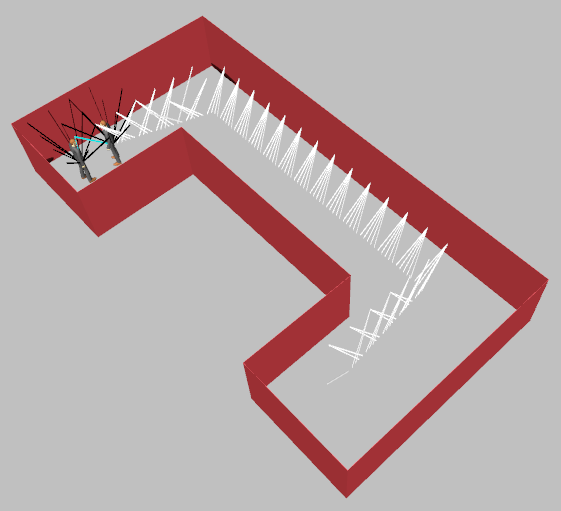}
\par\end{centering}
\caption{\label{fig:PropDoppler}Selected user links for Doppler effect mitigation
across a trajectory. }
\end{figure}
We proceed to examine the mitigation of Doppler Effects, using the
setup described in Table~\ref{tab:SetupDoppler}. Both users have
their antennas pointing upwards (i.e., $\phi=90^{o}$). User~$0$
moves along a trajectory, and $\text{\textsc{KpConfig}}$ proceeds
to filter its user links, selecting and employing those that are most
perpendicular to the trajectory. One or more paths can be established
to connect the pair, using one or more of the user~$0$ links. We
are interested in keeping the maximal deviation from perpendicularity
bounded; the maximal angle formed by each user link and the trajectory
must not exceed $90^{o}\pm10^{o}$. If no user link fulfills this
criterion, the link with the minimal deviation is selected.

\begin{figure}[t]
\begin{centering}
\subfloat[\label{fig:Trajectory-to-propagation-path-p}Trajectory-to-propagation
path perpendicularity.]{\begin{centering}
\includegraphics[viewport=0bp 0bp 425bp 150bp,clip,width=1\columnwidth]{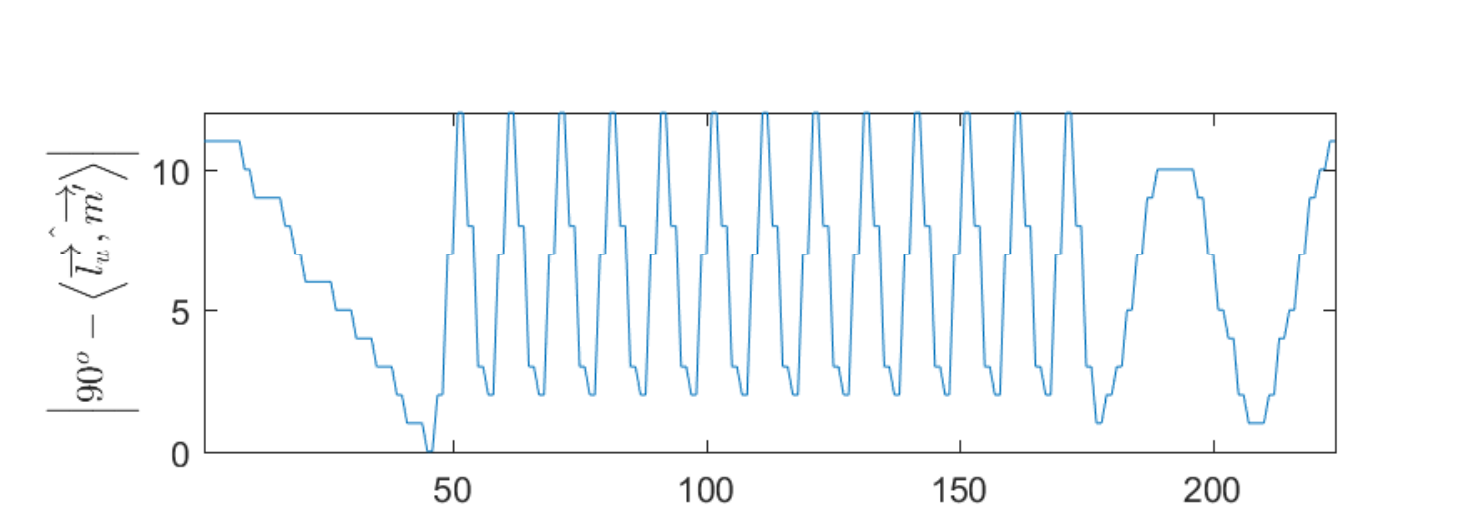}
\par\end{centering}
}
\par\end{centering}
\begin{centering}
\subfloat[\label{fig:Received-power.}Received power.]{\begin{centering}
\includegraphics[viewport=0bp 0bp 425bp 150bp,clip,width=1\columnwidth]{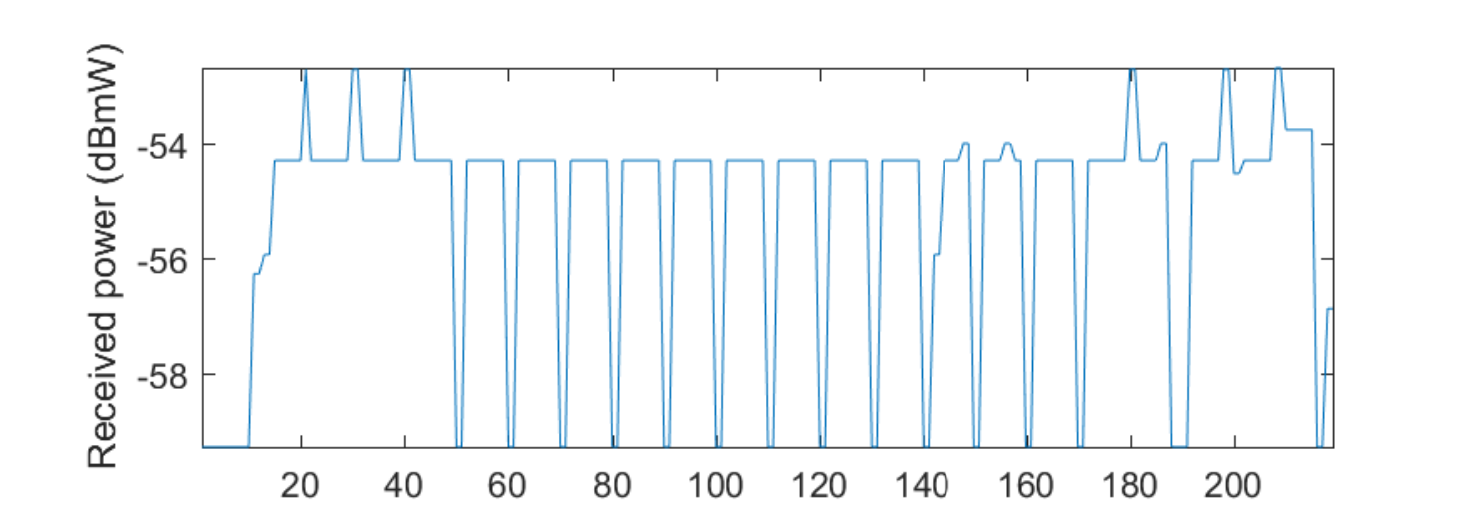}
\par\end{centering}
}
\par\end{centering}
\caption{\label{fig:RESULTS_doppler}Trajectory-to-propagation path perpendicularity
and receiver power across the user trajectory. (The x-axis represents
steps of \textasciitilde$12.5$ ~cm across the trajectory).}
\end{figure}

The user links selected across the trajectory are illustrated in Fig.~\ref{fig:PropDoppler}
as white lines. As the user moves, he is served by the tile nearest
to him. This tile remains active for a few steps and is then succeeded
by another one. When a user is exactly below each active tile, its
deviation from perpendicularity becomes minimal. The deviation then
increases until the user is positioned exactly below the next active
tile. This effect is shown in Fig.~\ref{fig:Trajectory-to-propagation-path-p},
which also showcases the effect of tile placement across the trajectory.
For the first $50$ steps, the user follows a skewed direction with
regard to the regular grid of the ceiling tiles (cf. the inset of
Table~\ref{tab:SetupDoppler}). As such, the tile selection and deviation
does not exhibit a well-defined pattern. For steps $50-175$, the
trajectory is aligned with tile centers at the ceiling, exhibiting
the described periodicity in the perpendicularity deviation. The pattern
changes again for the last, skewed part of the trajectory.

Figure~\ref{fig:Received-power.} shows the received power across
the trajectory. It is worth noting that the received power drops when
the deviation from perpendicularity spikes. Due to the user link selection
process described above, only one connecting path may be active in
these cases. This naturally limits the received power as well. Intermediate
points in the trajectory allow for 2 or 3 concurrent paths, increasing
the total carried power.

\begin{figure}[t]
\begin{centering}
\includegraphics[viewport=0bp 0bp 600bp 200bp,clip,width=1\columnwidth]{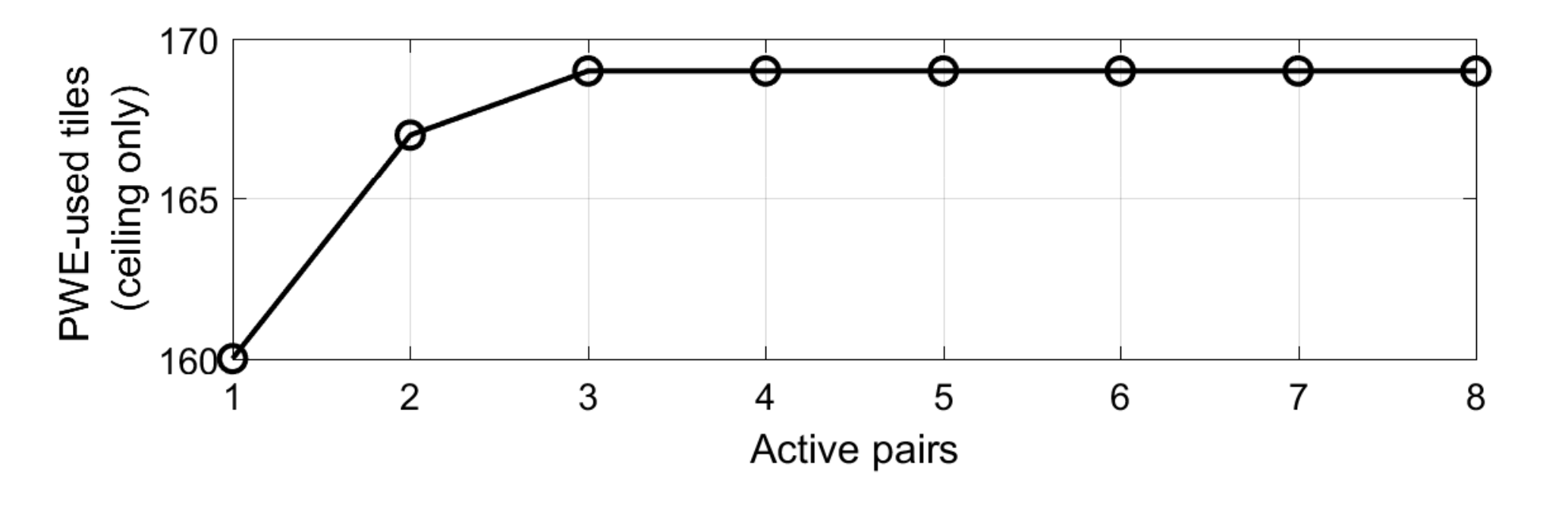}
\par\end{centering}
\caption{\label{fig:Tiles-activated-(out}Tiles activated (out of $169$ total)
as the number of user pairs increases, corresponding to the case of
Fig.~\ref{fig:S3}.}
\end{figure}
Finally, notice that the upwards directivity of the antennas means
that there is no LOS component across the trajectory. Moreover, the
use of collimation tile functions means that power losses are mainly
encountered over the user links (cf.~Fig.~\ref{fig3}). However,
the user link effects are static in this scenario, due to the upwards
directivity of the antennas. Thus, the received power is mainly affected
by the paths deployed between users $0$ and~$1$.

\subsection{User Capacity and Stress test}

\begin{table}[t]
\centering{}\caption{\label{tab:SetupStressTest}Scenario setup for the PWE stress test.}
\begin{tabular}[b]{cc}
\begin{tabular*}{3.8cm}{@{\extracolsep{\fill}}@{}l}
\toprule
\multicolumn{1}{l}{\textsc{User: Position, $\alpha$,$\phi$}}\tabularnewline
\midrule
0: {[}2.5,10.0,1.0{]},$\left\{ 50^{o},80^{o}\right\} $,90.0$^{o}$ \tabularnewline
1: {[}5.0,10.0,1.0{]},$\left\{ 50^{o},80^{o}\right\} $,90.0$^{o}$ \tabularnewline
2: {[}7.5,10.0,1.0{]},$\left\{ 50^{o},80^{o}\right\} $,90.0$^{o}$\tabularnewline
3: {[}10.0,10.0,1.0{]},$\left\{ 50^{o},80^{o}\right\} $,90.0$^{o}$\tabularnewline
4: {[}2.5,7.5,1.0{]},$\left\{ 50^{o},80^{o}\right\} $,90.0$^{o}$\tabularnewline
5: {[}5.0,7.5,1.0{]},$\left\{ 50^{o},80^{o}\right\} $,90.0$^{o}$\tabularnewline
6: {[}7.5,7.5,1.0{]},$\left\{ 50^{o},80^{o}\right\} $,90.0$^{o}$\tabularnewline
7: {[}10.0,7.5,1.0{]},$\left\{ 50^{o},80^{o}\right\} $,90.0$^{o}$\tabularnewline
8: {[}2.5,5.0,1.0{]},180.0$^{o}$,90.0$^{o}$\tabularnewline
9: {[}5.0,5.0,1.0{]},180.0$^{o}$,90.0$^{o}$\tabularnewline
10: {[}7.5,5.0,1.0{]},180.0$^{o}$,90.0$^{o}$\tabularnewline
11: {[}10.0,5.0,1.0{]},180.0$^{o}$,90.0$^{o}$\tabularnewline
12: {[}2.5,2.5,1.0{]},180.0$^{o}$,90.0$^{o}$\tabularnewline
13: {[}5.0,2.5,1.0{]},180.0$^{o}$,90.0$^{o}$\tabularnewline
14: {[}7.5,2.5,1.0{]},180.0$^{o}$,90.0$^{o}$\tabularnewline
15: {[}10.0,2.5,1.0{]},180.0$^{o}$,90.0$^{o}$\tabularnewline
\midrule
\multicolumn{1}{l}{\textsc{Pair: Objective}}\tabularnewline
\midrule
\textcolor{black}{$\left(i\right)\to\left(15-i\right),\,i=0\ldots7$}
: \tabularnewline
~~~~~~~${\scriptstyle {\textsc{MaxPower},\textsc{EavMit}[[\text{\textsc{All}}]]}}$\tabularnewline
\bottomrule
\end{tabular*} & %
\begin{tabular}{c}
\tabularnewline
\includegraphics[width=0.3\columnwidth]{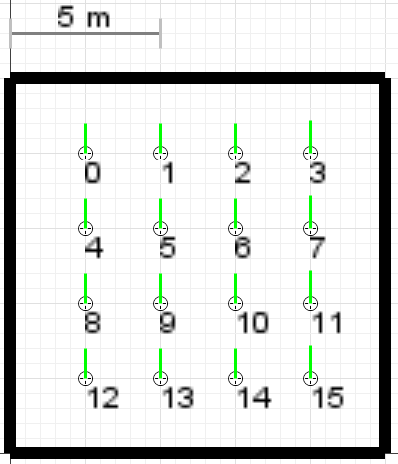}\tabularnewline
\end{tabular}\tabularnewline
\end{tabular}
\end{table}
\begin{figure}[!t]
\begin{centering}
\subfloat[\label{fig:S1}Full tile coverage, $\alpha=50^{o}$ for each transmitter.]{\begin{centering}
\includegraphics[viewport=100bp 240bp 500bp 550bp,clip,width=1\columnwidth]{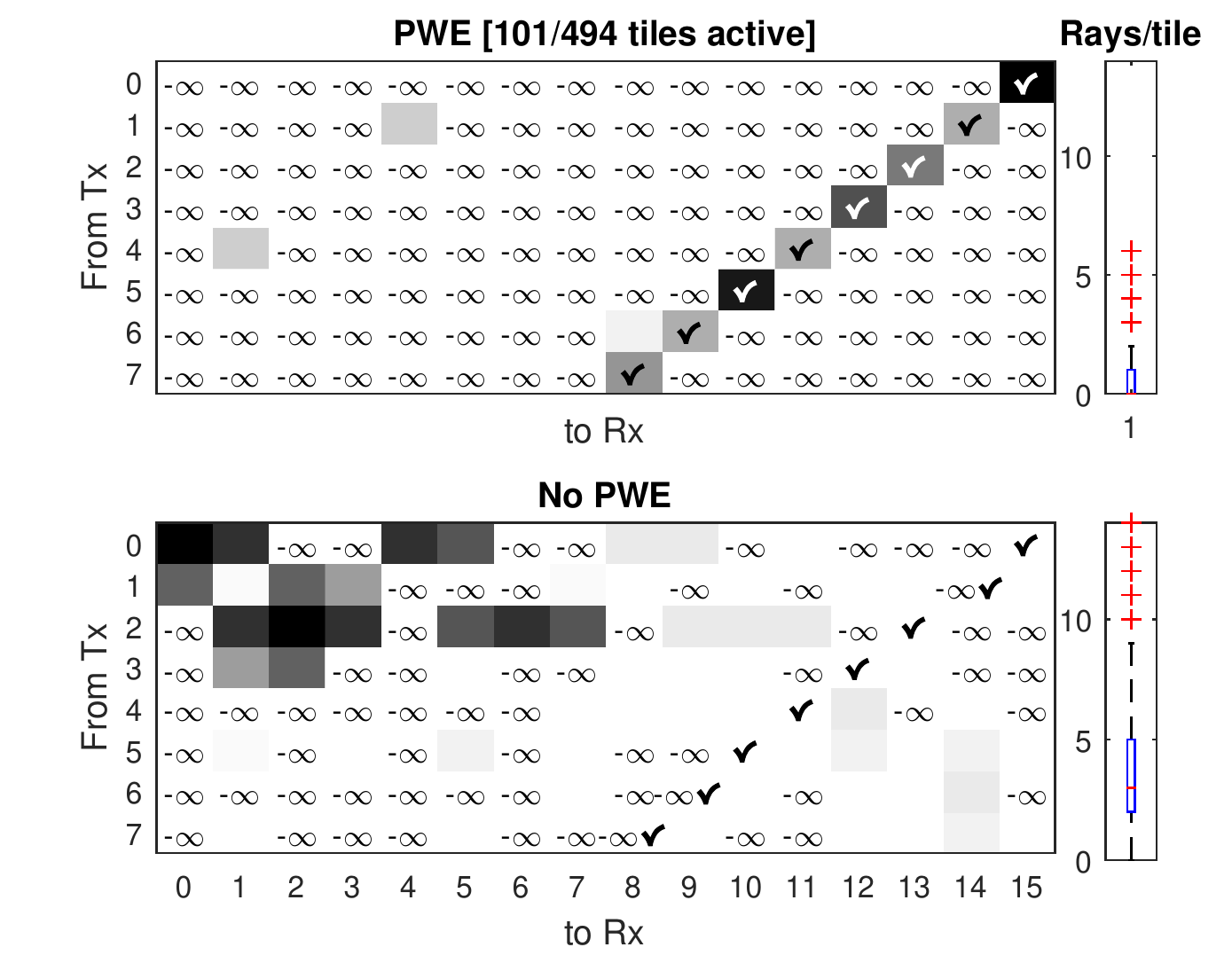}
\par\end{centering}
}
\par\end{centering}
\begin{centering}
\subfloat[\label{fig:S2}Full tile coverage, $\alpha=80^{o}$ for each transmitter.]{\begin{centering}
\includegraphics[viewport=100bp 240bp 500bp 550bp,clip,width=1\columnwidth]{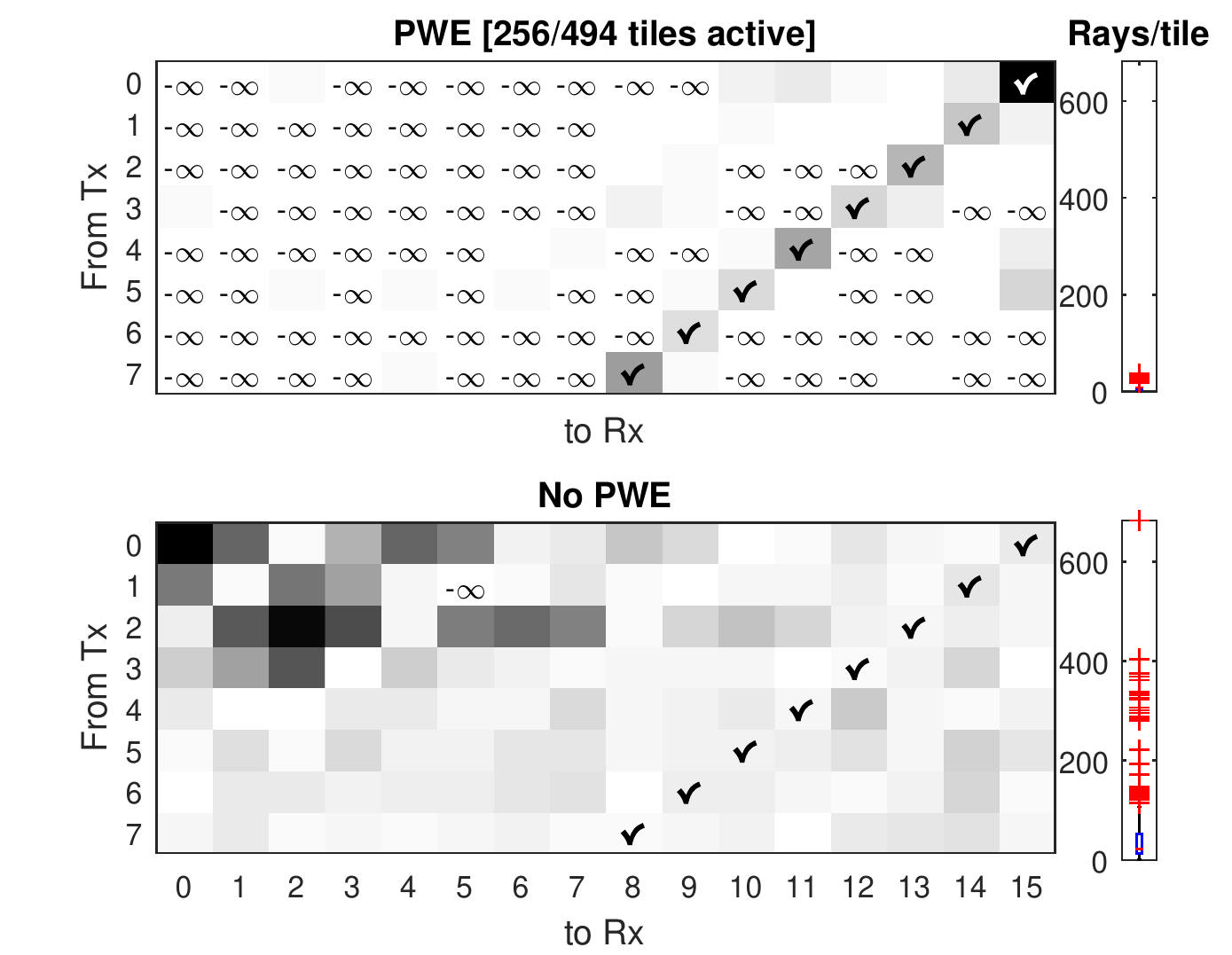}
\par\end{centering}
}
\par\end{centering}
\begin{centering}
\subfloat[\label{fig:S3}Partial tile coverage (ceiling only), $\alpha=80^{o}$
for each transmitter.]{\begin{centering}
\includegraphics[viewport=100bp 240bp 500bp 550bp,clip,width=1\columnwidth]{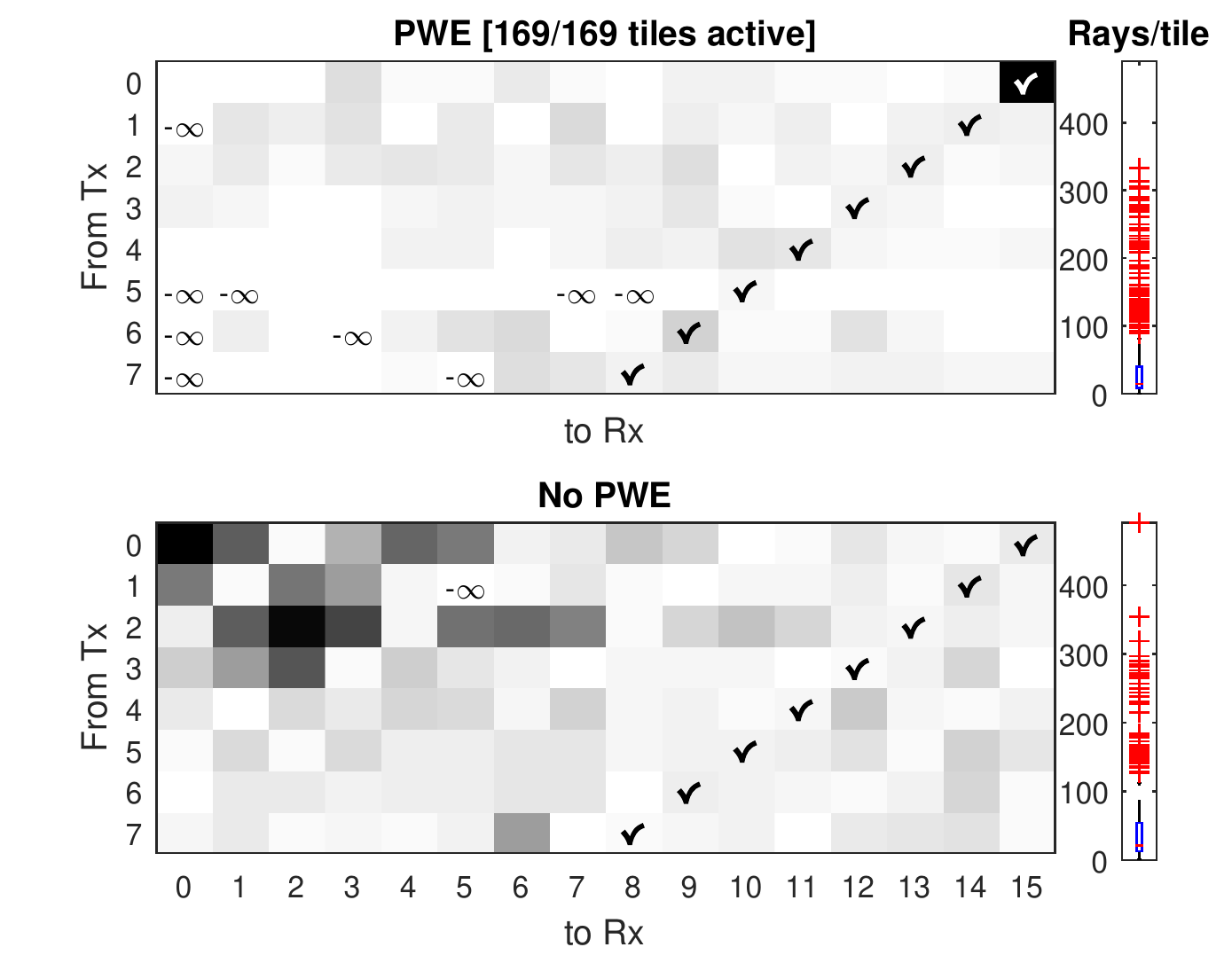}
\par\end{centering}
}
\par\end{centering}
\caption{\label{fig:RESULTSStressTest}Power reception and tile reuse for various
test of progressively increasing stress. \CheckmarkBold{} denotes
intended pair connectivity. The colormap is common across all subplots:
$\left[-106.94\right.$~~\protect\includegraphics[width=3cm,height=0.2cm]{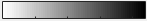}~~$\left.-59.23\right]$~dBmW.}
\end{figure}
The final test studies the capacity of a PWE, subject to a configuration
algorithm such as $\text{\textsc{KpConfig}}$. In other words, we
seek to understand the conditions under which the PWE performance
gains diminish to non-PWE levels.

The scenario setup is given in Table~\ref{tab:SetupStressTest}.
It comprises $8$ communicating pairs and three system stress levels.
In the first level all transmitters have antennas with $\alpha=50^{o}$,
while the HyperSurface coating is full over walls, floor and ceiling
($494$ tiles total). The second stress level is the same but with
$\alpha=80^{o}$, naturally leading to more emitted rays to handle
per transmitter. The last stress level is similar to the second, but
only the ceiling is HyperSurface-coated ($169$ tiles total). Thus,
the ray control points are fewer. Notice that partial coatings can
be accounted for in $\text{\textsc{KpConfig}}$ by passing a graph
input $\mathcal{G}$ that already contains tiles with $f_{h}\ne\emptyset$.
In other words, areas without tile coating are marked as having virtual
tiles, pre-configured with functions describing natural propagation.
Finally, all user antennas are pointing upwards, yielding only NLOS
components for all pairs.

Figure~\ref{fig:RESULTSStressTest} presents the received power and
interference among users per stress level, as well as the number of
rays impinging per tile in boxplot form. (All tiles are considered,
virtual or real, active or inactive). In the first stress level, the
PWE is configured by the $\text{\textsc{KpConfig}}$ almost ideally
(cf.~Fig.~\ref{fig:S1}). The intended pairs are connected and well-separated.
Minor interference is noticed explained as follows: with $\alpha=50^{o}$,
the affected tiles per transmitter marginally overlap. $\text{\textsc{KpConfig}}$
configures roughly $20\%$ of the available tiles ($\nicefrac{101}{494}$).
The tile re-use remains at very low levels, with $1$~ray per tile
being the most common case, as intended. The non-PWE does not exhibit
a well-defined pattern and yields approximately $30$~dBmW of extra
loss per pair, and complete disconnection for $3$ pairs.

The second stress level follows the same pattern but with decreased
performance in both PWE and non-PWE (Fig.~\ref{fig:S2}). The PWE
case continues to show good pair connectivity (notice the diagonal
of intended pairs), but interference has increased as expected. $\text{\textsc{KpConfig}}$
configures almost $50\%$ of the available tiles ($\nicefrac{256}{494}$).
The same effect is evident in the non-PWE case, validating the presence
of increased system stress. The tile re-use is kept low in the PWE
case.

The final stress case shows a collapse in the PWE performance, making
it almost inextinguishable from the non-PWE case (Fig.~\ref{fig:S3}).
$\text{\textsc{KpConfig}}$ uses all available tiles ($\nicefrac{169}{169}$).
The high stress level is also evident from the tile reuse, which is
almost identical for both PWE and non-PWE. In Figure~\ref{fig:Tiles-activated-(out}
we execute the same measurements as in Fig.~\ref{fig:RESULTSStressTest},
but by activating the transmitters in a serial fashion. Evidently,
after a single transmitter is activated, almost all of the PWE tiles
are used. This is natural, given that an antenna lobe of $\alpha=80^{o}$
pointing upwards affects almost all of the ceiling tiles. Subsequent
transmitters find no available free tiles to handle them, leading
to uncontrolled propagation. Thus, at this point the user capacity
of the PWE, \emph{subject to $\text{\textsc{KpConfig}}$ as the configuration
scheme}, has been exceeded.

\section{Discussion and Future Work\label{sec:Discussion-and-Future}}

The proposed \noun{KpConfig} algorithm constitutes a first of its
kind approach for configuring PWEs, with multi-user support and versatility
in expressing resource sharing policies and user communication objectives.
Moreover, it allows for expressing aspects of Physics in the form
of graph models and algorithms. Thus, it can provide a new field of
application for computer science concepts. For instance, subsequent
studies can study the relation between the form of PWE graphs and
the user capacity, as well as propose policies and alternative schemes
that behave better under stress.

While \noun{KpConfig} was employed in the preceding Sections for PWEs
that were either fully or partially coated with HyperSurface tiles.
The partially coating was arbitrarily set, to evaluate PWEs in a challenged
setup. However, partial coatings have the benefits of reduced cost
and easier deployment. In this aspect, deducing the partial PWE coatings
that yield maximal performance gains constitute a promising approach
for future research.

Handling transience, such as user mobility, was accomplished by executing
\noun{KpConfig} each time a change is detected in the environment.
Nonetheless, future extensions can take into account the current PWE
configuration, in order to minimize the changes in tile configurations.
For instance, minor changes in the location of a user may be accommodated
by modifying the last link of a path, rather than computing it from
scratch.

Additionally, as noted in preceding Sections, a PWE can only control
the NLOS component of communication, thus yielding a partial control
over propagation. Nonetheless, the control can be made full, assuming
joint collaboration with beamforming capabilities at the user devices.
As shown in Section~\ref{sec:Evaluation}, a device may seek to always
emit towards a HyperSurface-coated area first, thus eliminating the
LOS component and allowing for fully deterministic propagation in
total. Notice that NLOS in PWEs can yield exceptional performance.
Floorplan ceilings constitute good candidate-surfaces for applying
HyperSurface coatings given that: i) they are largely unused, ii)
they rarely contain obstacles, iii) they provide easy access to power
supply via the existing power lines for lights, and iv) the pose strong
security and interference mitigation benefits~\cite{Liaskos2019ADHOC}.
Thus, when present within a PWE, device gyroscopes can detect the
upwards directions and beamform for emissions accordingly.

Finally, it is noted that \noun{KpConfig} operates on the premise
of limiting tile re-use among users and function combinations in general.
However, future HyperSurface hardware technologies may provide perfect
performance for combined functions. In such scenarios, the core PWE
objective can be to minimize the number of activated tiles, thereby
favoring tile re-use.

\section{Conclusion\label{sec:Conclusion}}

Programmable wireless environments enable the customization of wireless
propagation within them. The present work presented a novel scheme\textendash the
\emph{$\text{\textsc{KpConfig}}$}\textendash to configure such environments
for serving multiple users and multiple objectives. Exemplary objectives
include security against eavesdropping, mitigating Doppler effects,
power transfer and signal-to-interference maximization. \emph{$\text{\textsc{KpConfig}}$}
employs a novel, graph-based model of programmable environments, which
transforms performance objectives to path search problems, while taking
into account core Physical restrictions. \emph{$\text{\textsc{KpConfig}}$
}was extensively evaluated in a novel tool for simulating programmable
wireless environments, yielding significant performance gains over
regular propagation and reaching insights on the user capacity of
PWEs.


\end{document}